\documentclass[aps,prc,amsmath,superscriptaddress,amssymb,twocolumn,showpacs]{revtex4-1} 
\usepackage{graphicx,amsmath,amssymb,bm,soul,cancel}

\usepackage[usenames]{color}

\newcommand{\ctu}[1]{\hat{c}_{\mathbf{#1} \uparrow}}

\newcommand{\mctd}[1]{\hat{c}_{-\mathbf{#1} \downarrow}}

\newcommand{\cts}[1]{\hat{c}_{\mathbf{#1} \sigma}}

\newcommand{\duo}[1]{\hat{p}_{\mathbf{#1}}}
\newcommand{\duod}[1]{\hat{p}_{\mathbf{#1}} ^\+}

\newcommand{\bra}[1]{\left< #1 \right|}
\newcommand{\ket}[1]{\left| #1 \right>}
\newcommand{\mv}[1]{\left< #1 \right>}
\newcommand{\comm}[2]{[ #1,#2 ]}

\newcommand{\mf}[1]{\mathbf{#1}}

\newcommand{\+}{\dagger}

\newcommand{\nks}[2]{\hat{n}_{\mf{#1} #2}}

\newcommand{\mfk}{\mf{k}}

\newcommand{\vn}{{\boldsymbol{\nu}}}

\newcommand{\vne}{{\hat{\nu}}}

\newcommand{\branch}[4]{
\left\{
	\begin{array}{ll}
		#1  & ,\,\,\, #2 \\ [0.3cm]
		#3  & ,\,\,\, #4
	\end{array}
	\right.
	}
\allowdisplaybreaks

\begin{document} 
 
 \title{From odd-even staggering to the pairing gap in neutron matter}

\author{Georgios Palkanoglou}
\affiliation{Department of Physics, University of Guelph, 
Guelph, ON N1G 2W1, Canada}
\author{Fotis~K.~Diakonos}
\affiliation{Department of Physics, University of Athens, 15771 Athens, Greece}
\author{Alexandros Gezerlis}
\affiliation{Department of Physics, University of Guelph, 
Guelph, ON N1G 2W1, Canada}

\begin{abstract} 
The properties of neutron matter are integral to the correct description of neutron stars as well as the description of neutron-rich nuclei. One key property of neutron matter is its superfluid behaviour in a range of densities relevant to the inner crust of neutron stars. This work investigates the finite-size effects in the pairing gap of a pure neutron matter superfluid system at densities found in the inner crust of cold neutron stars. The BCS (Bardeen-Cooper-Schrieffer) treatment of superfluidity gives rise to the mean-field pairing gap while a projection after variation (PBCS Theory) can provide a beyond-mean-field pairing gap through an odd-even staggering formula. While these two pairing gap results should agree in the thermodynamic limit, in this work we demonstrate that this is the case for systems far from the thermodynamic limit as well. These results can prove valuable to the study of neutron matter since they can connect \textit{ab initio} approaches to other approaches found in the literature. This is a first step towards a model-independent extraction of the pairing gap in neutron matter.
\end{abstract} 

%introduction
\maketitle 
\section{Introduction}
The physics of neutron-rich nuclei is connected to that of low-density neutron matter (NM)~\cite{Dean:2003, Gandolfi:2015} while properties of neutron stars (NSs) such as their cooling \cite{Yakovlev:2004, Page:2004, Page:2011} and the irregularities of their periods~\cite{Baym:1969, Watanabe:2017} can be traced back to interesting physics of the NM that makes up their inner and outer crusts. Furthermore, the equation of state (EoS) of high-density NM is integral to the determination of NS properties such as the mass-radius relation and the NS maximum mass, while accurately calculated properties of NM can be used to provide a benchmark for nuclear energy-density functionals (EDFs) \cite{Gezerlis:book:2014} which in turn can be used to guide hydrodynamic descriptions of the inner crust of NSs \cite{Chamel:2017}. Finally, one can find a correspondence between NM and cold atoms: unitarity is the regime of a Fermi gas with $k_\textrm{F} a\to -\infty$ where $a$ is the scattering length, and negligible effective range $k_\textrm{F} r_e \approx 0$. In such systems all length scales drop out of the problem and one expects a universal behaviour~\cite{Bertsch:2001}. This regime is located on the BCS-BEC crossover which is encountered by increasing $1/k_\textrm{F}a$ from negative to positive (from $k_\textrm{F} a\to -\infty$ to $k_\textrm{F} a\to \infty$) where $k_\textrm{F}$ is the Fermi momentum. This crossover can be generalized for finite $r_\textrm{e}$~\cite{Tajima:2019} to $-1/(k_\textrm{F}a) + r_\textrm{e}k_\textrm{F}/2 = 0$. The bare neutron-neutron (NN) interaction is characterized by a very large scattering length $(a\approx-18.5~\textrm{fm}$) which means that, to the extent that the finite effective range of the NN interaction can be neglected ($r_e \approx 2.7~\textrm{fm}$), the properties of a dilute neutron gas can be considered close to that of a unitary gas. This allows a connection between NM and cold atoms close to unitarity which has motivated and benefited by various theoretical~\cite{Strinati:2018, Baker:1999, deMelo:1993, Heiselberg:2001, Gezerlis:2008, Carlson:2003} and experimental studies~\cite{Bourdel:2003, Regal:2003}.

Pairing in nuclear systems has been a longstanding area of research for the past half a century. In almost all known nuclei, one can find isovector ($T=1$, $S=0$) neutron-neutron ($nn$) and proton-proton ($pp$) pairing dominating. In $N \approx Z$ nuclei, isovector and isoscalar ($T=0$, $S=1$) neutron-proton ($np$) pairing should be present with the latter being notoriously elusive~\cite{Frauendorf:2014, Baroni:2010, Romero:2019}. The inner crust of a cold NS consists of a fluid of neutrons permeating a crystal lattice of heavy (neutron-rich) nuclei. The density of these neutrons is slightly less than the nuclear saturation density $n_0 = 0.16~\textrm{fm}^{-3}$. At low densities the NN interaction is attractive mainly through the $^1S_0$ channel causing the creation of neutron isovector pairing which in turn brings the neutron fluid to a superfluid state~\cite{Dean:2003}. Deeper in the crust isovector proton pairing is also present~\cite{Gezerlis:book:2014}. As the density increases by depth, the NN interaction becomes repulsive in the $^1S_0$ channel closing superfluidity through that channel at densities that correspond to $k_{\textrm{F}} \approx 1.5~\textrm{fm}^{-1}$. From that point on, the dominant component of attractive interactions comes from the $^3P_2$ channel which is coupled to the $^3F_2$ channel. This $p$-wave attractive interaction has been shown to be crucial to the description of the NS structure at higher densities since it corrects the instability due to the the repulsiveness of the $s$-wave interaction at these densities~\cite{Bonnard:2020}. The exponentially suppressed heat capacity of the superfluid state and the scattering of electrons in the ``normal state" cores of the vortex lines in the superfluid inner crust impact the observed cooling of NSs~\cite{Yakovlev:2004, Page:2004, Page:2011} and the glitches of pulsars~\cite{Baym:1969}, respectively. A correct description of neutron pairing is important for the understanding of such phenomena (see Ref.~\cite{Gezerlis:book:2014} for a review on the superfluidity in NSs).

Calculations of the pairing gap in NM have been done in the past for the $^1S_0$ pairing gap for realistic interactions set to reproduce the scattering length and effective range of the bare NN interaction. Such studies have been conducted in the BCS framework~\cite{Dean:2003, Gezerlis:2008} and beyond by the inclusion of short and long range correlations in the gap equations~\cite{Ding:2016} or by the means of correlated basis functions (CBF)~\cite{Pavlou:2017} where one describes the ground state of the system with the use of correlation operators~\cite{Kortscheck:1980, Fan:2018}. Calculations of the $^1S_0$ NM pairing gap have also been done employing interactions tuned to reproduce other well-established physics of NM~\cite{Chamel:2008} as well as chiral interactions~\cite{Hebeler:2007, Hebeler:2010, Maurizio:2014, Ding:2016}. General studies of strongly paired fermions, of which NM is a subcategory, have been also conducted using effective field theory (see \cite{Hu:2020} and references therein). These calculations refer to a pairing gap that is defined as the minimum of the corresponding quasi-particle excitation energy. Finally, to these one should add \textit{ab initio} calculations of the $^1S_0$ pairing gap which have been done using Quantum Monte Carlo (QMC) techniques~\cite{Gezerlis:2010, Carlson:2003, Carlson:2005} for finite particle numbers and then extrapolated to the thermodynamic limit (TL). These techniques utilize an odd-even staggering (OES) definition where the pairing gap is computed as the energy difference between systems with fully paired particles and systems with one unpaired particle. While these two definitions are equivalent at the TL, the relationship between the two far from the TL is not trivial. Aiming to bridge this gap, we performed mean-field calculations of the pairing gap in finite superfluid systems which then were extended to a beyond-mean-field approach by the means of symmetry restoration techniques. The pairing gaps resulting from the two approaches were then compared far from the TL where they were found to agree with each other. In the context of \textit{ab initio} approaches, where one faces the task of extrapolating studies of finite systems to the TL \cite{Gezerlis:2008,Jensen:2020}(for higher densities see Refs~\cite{Hagen:2013, Gezerlis:2014}), these results can be used to connect studies of superfluid systems to other approaches in the literature. Thus, this is a first step towards a model-independent extraction of the pairing gap in neutron matter.

\section{The BCS Theory for Neutron Matter}
\label{sec:BCS}
We investigate the effects of pairing in the NM found in the inner crust of NSs. At first order, the NM of the inner crust of a cold NS can be approximated by infinite pure NM. On a mean-field level one can use the bare NN interaction within the BCS theory of superconductivity to describe the NM pairing correlations~\cite{Gezerlis:2010}. In more sophisticated approaches one could consider induced interactions stemming from the small component of protons in the inner crust as well as screening and anti-screening effects~\cite{Ramanan:2018}, both of which are beyond the scope of this work. We are interested in the properties of the bulk medium in pure NM and, therefore, we enclose the system in a box of length $L$, much larger than the effective range of the NN interaction, employing periodic boundary conditions (PBC). The choice of PBC comes naturally from the observation that in uniform infinite matter all physical properties must be invariant under spatial translation.

\subsection{Even-particle-number systems}
\label{sec:BCS:sub:Even}
According to the BCS theory, the normal state of a fluid of unpolarized fermions (half of which are spin-up and half spin-down) exhibits an instability in the presence of attractive interactions. In the formulation of the theory, the ground state of the system is described as a superposition of pairs of time-reversed states:

\begin{align}
\left| \psi _{\textrm{BCS}} \right> = \prod_{\mathbf{k}} \left(u_{\mathbf{k}} + v_{\mathbf{k}} \hat{c}_{\mathbf{k}\uparrow}^{\dagger} \hat{c}_{-\mathbf{k}\downarrow}^{\dagger}\right)\left|0\right> ~,\label{eq:groundstateBCS}
\end{align}

\noindent where $\hat{c}_{\mathbf{k} \sigma}^{\dagger}, \hat{c}_{\mathbf{k} \sigma}$ are fermionic creation and annihilation operators, respectively, that are associated with the single-particle wave functions of particles of momentum $\mathbf{k}$ and spin $\sigma$ in a cubic box under PBC and $\left|0\right>$ is the vacuum state. The state in Eq.~(\ref{eq:groundstateBCS}) describes systems with even particle-numbers while, as will be seen in Subsection \ref{sec:BCS:sub:Odd}, a minor modification can be done in Eq.~(\ref{eq:groundstateBCS}) to describe systems with odd particle-numbers. The functions $v_{\mathbf{k}}^2$ and $u_{\mathbf{k}}^2$ are the probabilities of finding or not finding, respectively, a pair with momenta $\mathbf{k}\uparrow, -\mathbf{k}\downarrow$, and as such their sum for a given $\mathbf{k}$ is normalized to unity:
\begin{align}
v_{\mathbf{k}}^2 + u_{\mathbf{k}}^2 = 1 ~.\label{eq:constraint}
\end{align}

Our aim is to study the effects of pairing and as such we employ a Hamiltonian where one ignores normal state interactions, as is standard in the literature:  
\begin{align}
\hat{H} = \sum_{\mathbf{k} \sigma} \epsilon _\mathbf{k} \hat{c}_{\mathbf{k} \sigma}^\dagger \hat{c}_{\mathbf{k} \sigma} + \sum_{\mathbf{k}\mathbf{l}} \left<\mathbf{k}\right|V\left|\mathbf{l}\right> \hat{c}_{\mathbf{k} \uparrow}^\dagger\hat{c}_{-\mathbf{k} \downarrow}^\dagger\hat{c}_{-\mathbf{l} \downarrow}\hat{c}_{\mathbf{l} \uparrow} ~,\label{eq:Hamiltonian}
\end{align}
where $\epsilon _\mathbf{k}$ is the single-particle energy associated with the momentum state $\mathbf{k}$ and $\left<\mathbf{k}\right|V\left|\mathbf{l}\right>$ the matrix element of the pairing interaction, i.e., the attractive interaction responsible for the instability against the pairing. The state in Eq.~(\ref{eq:groundstateBCS}) is not an eigenstate of the number operator, $\hat{N} = \sum _{\mathbf{k} \sigma} \hat{c}_{\mathbf{k} \sigma}^\dagger \hat{c}_{\mathbf{k} \sigma}$. Because of that, BCS theory is formulated in a grand canonical ensemble such that the average number of particles remains fixed:
\begin{align}
\left<\hat{N}\right> &= \left< \psi _{\textrm{BCS}} \right|  \left(\sum _{\mathbf{k}} \hat{c}_{\mathbf{k} \uparrow}^\dagger \hat{c}_{\mathbf{k} \uparrow} + \hat{c}_{\mathbf{k} \downarrow}^\dagger \hat{c}_{\mathbf{k} \downarrow} \right)  \left| \psi _{\textrm{BCS}} \right> \nonumber \\
	 &= 2\sum _{\mathbf{k}} v_{\mathbf{k}}^2   ~. \label{eq:AvgParticleNumber}
\end{align}

\begin{figure}[htp]
\begin{center}
\includegraphics[width=1.0\columnwidth,clip=]{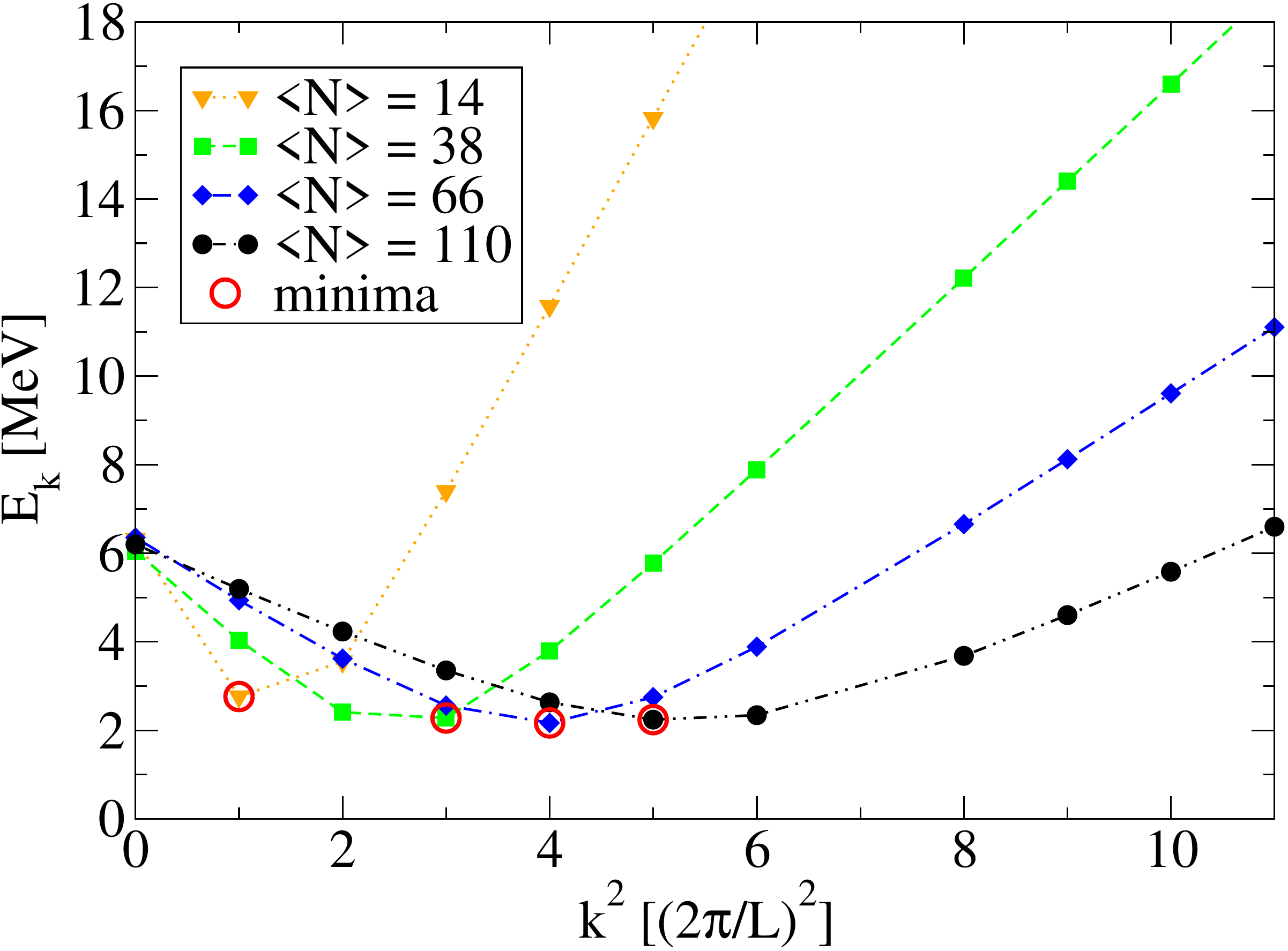}
\caption{The quasi-particle excitation energy, at $k_{\textrm{F}}a=-10$, as a function of the square of the momentum magnitude $k^2$ for a set of different average particle numbers $\left<N\right>$ chosen to clearly demonstrate the shift of the minimum of the quasi-particle excitation energy as a function of $\left<N\right>$. Circled on the figure are the positions of the minima of the excitation energies for different $\left<N\right>$ . The value of these minima correspond to the mean-field pairing gap.	\label{fig:Exc}}
\end{center}
\end{figure}

The ground state of the system is determined through a variational approach: the distributions $v_\mathbf{k}$ and $u_\mathbf{k}$ are determined so that they minimize the energy of the state in Eq.~(\ref{eq:groundstateBCS}),
\begin{align}
\left<\psi _{\textrm{BCS}} \right| \hat{H} \left|\psi _{\textrm{BCS}}\right> & = \sum_{\mathbf{k} \sigma} \epsilon _{\mathbf{k}} v_\mathbf{k}^2 + \nonumber  \\	
	&+ \sum_{\mathbf{k}{\mathbf{k'}}} \left<\mathbf{k}\right|V\left|\mathbf{k'}\right>  u_\mathbf{k}v_\mathbf{k}u_{\mathbf{k'}} v_{\mathbf{k'}}   ~,	\label{eq:EnergyBCS}
\end{align}
while respecting the constraint in Eq.~(\ref{eq:AvgParticleNumber}). Using the Lagrange multiplier scheme, this minimization is equivalent to minimizing the quantity:
\pagebreak[4]
\begin{align}
W[v_{\mathbf{k}};\mu] &= \mv{\hat{H}} - \mu \left(\mv{\hat{N}} - N_0\right)  \nonumber\\
	&=\sum_{\mathbf{k} \sigma} \xi _{\mathbf{k}} v_\mathbf{k}^2 + \sum_{\mathbf{k}{\mathbf{k'}}} \left<\mathbf{k}\right|V\left|\mathbf{k'}\right>  u_\mathbf{k}v_\mathbf{k}u_{\mathbf{k'}} v_{\mathbf{k'}} \nonumber\\
		&+\mu N_0   ~,	\label{eq:FreeEnergy}
\end{align}
where $\xi _{\mathbf{k}} = \epsilon _{\mathbf{k}} - \mu$~\cite{deGennes:Book}. The single-particle energies $\epsilon_{\mathbf{k}}$ are:
\begin{align}
\epsilon _\mathbf{k} &= \frac{\hbar ^2}{2m} \left|\mathbf{k}\right| ^2~.
\end{align}
The quantity $N_0$ is the desired average particle number of the system and $\mu$ the chemical potential (the Lagrange multiplier). In Eq.~(\ref{eq:FreeEnergy}) we have neglected terms that come from the diagonal elements of the interaction matrix in Eq.~(\ref{eq:Hamiltonian}). These terms' only effect, when grouped with the kinetic term in Eq.~(\ref{eq:FreeEnergy}), is the renormalization of the single-particle energies~\cite{NuclMB:Book}~(p.~238). Note that the explicit dependence of $W$ on $u_\mathbf{k}$ was omitted since the distribution $u_\mathbf{k}$ can be uniquely defined through Eq.~(\ref{eq:constraint}) for a given distribution $v_\mathbf{k}$. Taking the variation of $W$ with respect to $v_\mathbf{k}$ we arrive at the famous BCS gap equation whose solution determines $v_{\mathbf{k}}^2$ and $u_{\mathbf{k}}^2$:
\begin{align}
\Delta _{\mathbf{k}} = -\frac{1}{2}\sum_{{\mathbf{k'}}} \left<\mathbf{k}\right|V\left|\mathbf{k'}\right>  \frac{\Delta _{\mathbf{k'}}}{E _{\mathbf{k'}}}   ~,\label{eq:Gap1BCS} 
\end{align}
where the gap function is:
\begin{align}
\Delta _{\mathbf{k}} = - \sum _{{\mathbf{k'}}} \left<\mathbf{k}\right|V\left|{\mathbf{k'}}\right> v_{{\mathbf{k'}}} u_{{\mathbf{k'}}} ~. \label{eq:Gappdefinition}
\end{align}
Here, $E_{\mathbf{k}}$ is the quasi-particle excitation energy, i.e., the energy needed to create an excitation by breaking a pair at a state $\mathbf{k}$, and it is defined as:
\begin{align}
E _{\mathbf{k}} = \sqrt{\xi _{\mathbf{k}}^2 + \Delta _{\mathbf{k}}^2}	\label{eq:ExcEnergydefinition} ~.
\end{align}
Solving Eqs.~{(\ref{eq:Gap1BCS})~{\&}~(\ref{eq:Gap2BCS})} one obtains the gap distribution $\Delta _{\mathbf{k}}$ and the chemical potential $\mu $ for a given average particle number $\left< N \right>$. A plot of the quasi-particle excitation energy using the solutions of the gap equations for finite systems with various particle numbers $\left< N\right>$ can be seen in Fig.~\ref{fig:Exc} as a function of the momentum squared. The minimum of the quasi-particle excitation energy is defined as the pairing gap:
\begin{align}
    \Delta _{\textrm{MF}} = \textrm{min}_{\mathbf{k}}E_{\mathbf{k}}~. \label{eq:D_MF}
\end{align}
As will be discussed in Sec.~\ref{sec:OES}, the pairing gap is a measure of the pairing correlation in the superfluid. The subscript MF above refers to the fact that the pairing gap calculated using Eq.~(\ref{eq:D_MF}) is a result of a pure mean-field treatment and as such contains no beyond-mean-field contributions.

One should note that we are describing an interacting system and as such the concepts of a Fermi energy $E_{\textrm{F}}$ and a Fermi momentum $k_{\textrm{F}}$ should be understood as an energy scale introduced by the density $n$ of the inner crust as:
\begin{align}
E_{\textrm{F}} = \frac{\hbar ^2}{2m} k_{\textrm{F}}^2 = \frac{\hbar ^2}{2m}\left(3\pi ^2n\right) ^{2/3} ~.
\end{align}
As our goal is to study the trend of a finite system towards the TL, we want to focus on intensive quantities of the system since these will remain finite at the TL. Therefore it is more suitable for one to focus on the energy per particle as opposed to the pure energy of the system. This introduces an additional energy constant which is the energy per particle of a free Fermi gas at the TL: 
\begin{align}
    \frac{E}{N}\bigg|_{\textrm{TL}} = \frac{3}{5}E_{\textrm{F}}
\end{align}

It should also be noted that hereinafter we will refer to the density using the dimensionless parameter $k_{\textrm{F}}a$ where $a \approx -18.5~{\textrm{fm}}$ is the scattering length of NM. The BCS formalism was first expressed in terms of the scattering length in Ref.~\cite{Legget}. Since then it has been customary to study the properties of superfluid dilute Fermi gases as a function of $k_{\textrm{F}}a$, first used in Ref.~\cite{deMelo:1993}.

In terms of the quasi-particle excitation energy and the gap distribution, the distributions $v_\mathbf{k}$ and $u_\mathbf{k}$ are:
\begin{align}
v_{\mathbf{k}}^2 &= \frac{1}{2}\left(1-\frac{\xi _{\mathbf{k}}}{E _{\mathbf{k}}}\right)  ~, \label{eq:v2kdefinition}\\ 
u_{\mathbf{k}}^2 &= \frac{1}{2}\left(1+\frac{\xi _{\mathbf{k}}}{E _{\mathbf{k}}}\right)  ~. \label{eq:u2kdefinition}
\end{align}
The above relations can be derived by solving for $v_\mathbf{k}$ (or $u_\mathbf{k}$) in $2u_\mathbf{k} v_\mathbf{k} = \Delta _\mathbf{k}/E_\mathbf{k}$ along with Eq.~(\ref{eq:constraint}). Taking the variation of $W$ with respect to the Lagrange multiplier $\mu$ we find:
\begin{align}
\left< \hat{N}\right> = \sum_{\mathbf{k}}\left(1-\frac{\xi _\mathbf{k}}{E _{\mathbf{k}}}\right) = N_0  ~,\label{eq:Gap2BCS} 
\end{align}

\noindent which is nothing more than the average particle-number conservation that one finds in the grand canonical ensemble. The distribution $v^2_\mathbf{k}$ is the probability of finding a pair at a momentum $\mathbf{k}$. A substantial smearing of that distribution over $\mathbf{k}$-space can be seen in Fig.~\ref{fig:v2k}. That smearing is proportional to the mean-field pairing gap and it is a consequence of strong pairing correlations. The specific particle numbers plotted both in that figure and Fig.~\ref{fig:Exc} are illustrative.

Equations~{(\ref{eq:Gap1BCS})~{\&}~(\ref{eq:Gap2BCS})} are two coupled non-linear equations in the sense that Eq.~(\ref{eq:Gap1BCS}) contains the gap distribution $\Delta _{\mathbf{k}}$ both on the LHS and the RHS in a non-separable way (non-linear) and both Eqs.~{(\ref{eq:Gap1BCS})~{\&}~(\ref{eq:Gap2BCS}) contain the unknowns $\Delta _{\mathbf{k}},\, \mu$ in such a way that one is unable to solve the one and substitute it into the other (coupled). Equations~{(\ref{eq:Gap1BCS})~{\&}~(\ref{eq:Gap2BCS})} are usually referred to as the BCS gap equations. They can be decoupled in the weak-coupling limit where $\Delta / \mu \ll 1$. That condition, however, is not met for NM and one is faced with the task of solving the BCS gap equations self-consistently. 

Before engaging in such a task, it has to be ensured that the matrix element in Eq.~(\ref{eq:Gap1BCS}) is the interaction responsible for the pairing. In NM at the densities considered here, this interaction comes mainly from the $^1S_0$~channel of the NN interaction. Therefore, the matrix element in Eq.~(\ref{eq:Gap1BCS}) must be expanded in partial waves where only the S-wave is to be kept. That leads to the angle-averaged version of the BCS gap equations:

\begin{align}
\Delta (k) &= -\frac{2\pi}{L^3} \sum_{k'} M(k') V_0(k,k')\frac{\Delta(k')}{E(k')} 	~, \label{eq:Gap1BCSSwave} \\
\left<\hat{N}\right> &= \sum _{k} M(k) \left(1-\frac{\xi(k)}{E(k)}\right)	 ~, 	\label{eq:Gap2BCSSwave}
\end{align}   
where $L$ is the length of the (cubic) box and $V_0(k,k')$ is the matrix element of the potential averaged over the angle between $\mathbf{k}$ and $\mathbf{k'}$:
\begin{align}
V_0(k,k') = \int _0^\infty dr\, r^2 j_0\left(kr\right)V(r)j_0\left(k'r\right) ~,	\label{eq:SwavePotential}
\end{align} 
\begin{figure}
\begin{center}
\includegraphics[width=1.0\columnwidth,clip=]{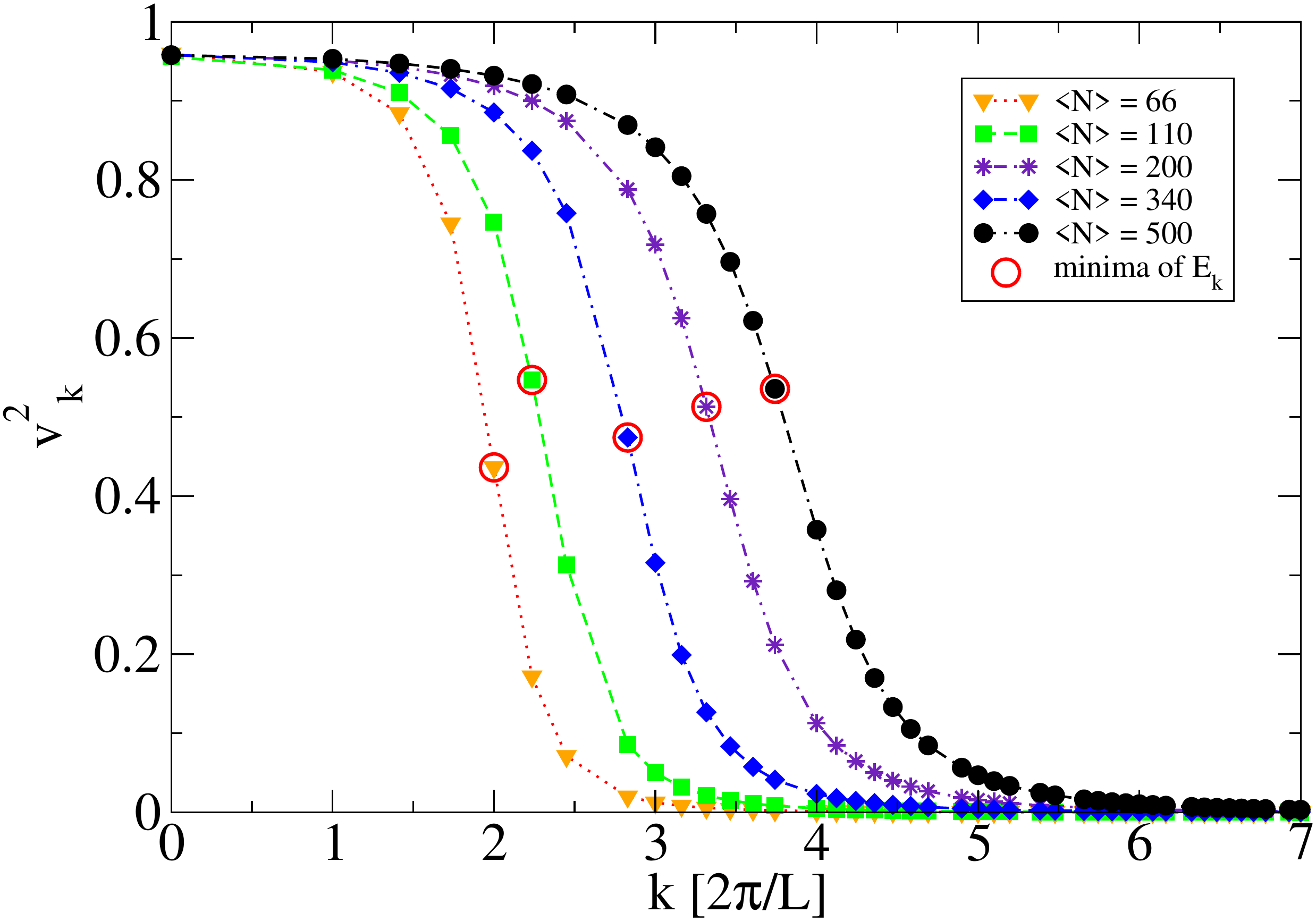}
\caption{The pair occupation probability, at $k_{\textrm{F}}a=-10$, as a function of the momentum magnitude $k$ for a set of different average particle-numbers $\left<N\right>$ chosen to clearly demonstrate the dependence of the occupation of the shell that corresponds to the minimum of $E_{\mf{k}}$ on the average particle-number.  Circled on the figure are the positions of the minima of the excitation energies for different $\left<N\right>$. A key feature of strong pairing is the smearing of the probability distributions.	\label{fig:v2k}}
\end{center}
\end{figure}
with $j_0(kr)$ being the zeroth-order spherical Bessel function of the first kind. Here $M(k')$ is the population function which counts the number of $\mathbf{k}$-states that correspond to magnitude $k'$. The full derivation of the above equations can be found in Ref.~\cite{Drischler:2017}. We present a simplified version in Appendix~\ref{App:SwaveGap}.

Within the BCS framework, the energy of even-particle-number systems is given by the ground state expectation value of the Hamiltonian, namely Eq.~(\ref{eq:EnergyBCS}). In a way similar to that of the BCS gap equations, the $S$-wave of the pairing interaction in this equations has to be isolated leading to the following equation for even-particle-number systems:
\begin{align}
E^{\textrm{BCS}}_{\textrm{even}} & (N) = \sum_{k} M(k) \epsilon _{k} 2v_k^2 + 	\nonumber \\
	&+ \frac{4\pi}{L^3} \sum_{k k'} M(k)M(k') V_0(k,k') u_kv_k u_{k'} v_{k'} ~ ,	\label{eq:SwaveEnergyBCS}
\end{align}
where M(k) is again the population function and the quantities $v_\mathbf{k}^2$ and $v_\mathbf{k} u_\mathbf{k}v_\mathbf{k'} u_\mathbf{k'}$ have been replaced by their angle-averaged counterparts. We have also used Eq.~(\ref{eq:D2.1}).

\subsection{Odd-particle-number systems}
\label{sec:BCS:sub:Odd}
As discussed above, the BCS ground state (see Eq.~(\ref{eq:groundstateBCS})) describes the condensate as a superposition of pair-states and as such it can only describe systems with an even number of particles. Odd-particle-number systems are described employing blocked states in which the extra, unpaired particle will occupy a momentum state $\mathbf{b}$ blocking the formation of a pair on it. Those systems, at the BCS level, are described by the state:
\begin{align}
\left|\psi _{\textrm{BCS}}^{\mathbf{b} \gamma}\right> = \hat{c}_{\mathbf{b}\gamma}^\dagger \prod_{\mathbf{k} \neq \mathbf{b}} \left(u_\mathbf{k} + v_\mathbf{k} \hat{c}_{\mathbf{k} \uparrow}^\+ \hat{c}_{-\mathbf{k} \downarrow}^\+\right)\left| 0\right> ~,	\label{eq:bBCS} 
\end{align}
\begin{figure}[htp]
\begin{center}
\includegraphics[width=1.0\columnwidth,clip=]{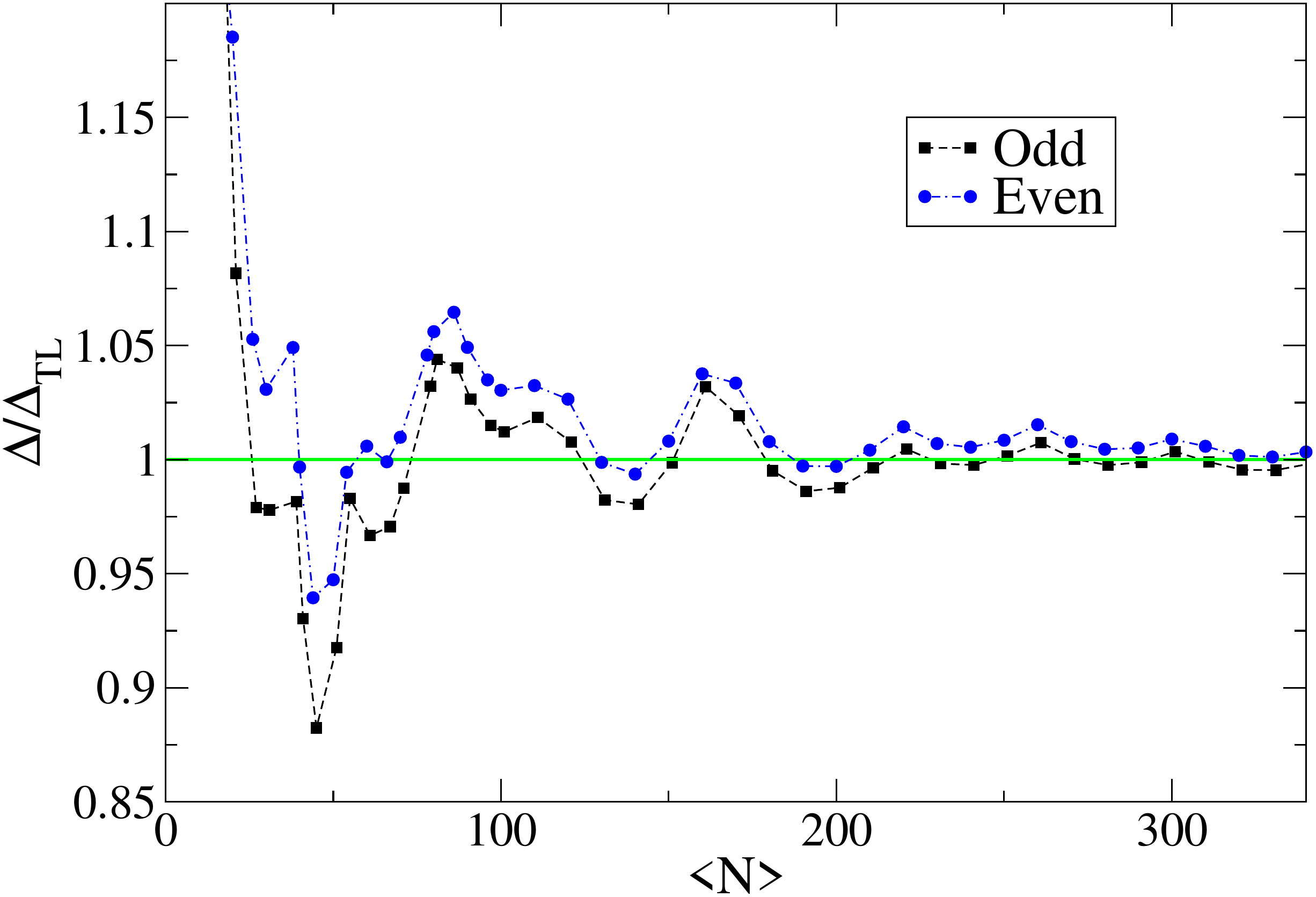}
\caption{The pairing gap as a minimum of the quasi-particle excitation energy in odd and even particle-number systems in BCS at $k_{\textrm{F}}a=-10$. \label{fig:N_D_MF}}
\end{center}
\end{figure}
where $\mathbf{b}$ and $\gamma$ are the momentum and the spin projection, respectively, of the unpaired particle. The distributions $v_\mathbf{k}$ and $u_\mathbf{k}$ come from this state's own self-consistent treatment: one needs to minimize the energy of the state subject to the constraint of the fixed average particle-number as in Eq.~(\ref{eq:FreeEnergy}). That results in the blocked BCS gap equations:
%\begin{align}
%\Delta (k) &= -\frac{2\pi}{L^3} \sum_{k'} \left(M(k')-\delta _{k'b}\right) V_0(k,k')\frac{\Delta(k')}{E(k')} 	 ~,\label{eq:bGap1BCSSwave} \\
%\left<\hat{N}\right> - 1 &= \sum _{k} \left(M(k)  - \delta _{kb}\right) \left(1-\frac{\xi(k)}{E(k)}\right)	~ ,	\label{eq:bGap2BCSSwave}
%\end{align}  
\begin{align}
\Delta (k) &= -\frac{2\pi}{L^3} \sum_{k'\neq b} M(k') V_0(k,k')\frac{\Delta(k')}{E(k')} 	 ~,\label{eq:bGap1BCSSwave} \\
\left<\hat{N}\right> - 1 &= \sum _{k\neq b} M(k) \left(1-\frac{\xi(k)}{E(k)}\right)	~ ,	\label{eq:bGap2BCSSwave}
\end{align}  
where we have again kept only the $S$-wave of the pairing interaction. We should note that there is a slight abuse of notation in Eqs.~(\ref{eq:bGap1BCSSwave})~\&~(\ref{eq:bGap2BCSSwave}): the symbol $\sum_{k\neq b}$ signifies the blocking of only one $\mathbf{k}$-state and not the entire shell of $\mathbf{k}$-states corresponding to $\left|\mathbf{k}\right|=k$. 
We see that the blocked state describes a superfluid of $N_0-1$ particles which has no access to the blocked state $\mathbf{b}$ and a particle on the blocked state $\mathbf{b}$ which is essentially a free particle and its effect on the condensate is only through the restriction that the blocking imposes on the available $\mathbf{k}$-space. This can be seen by inspecting Eq.~{(\ref{eq:Hamiltonian})} or more clearly in the energy that corresponds to the blocked state:
\begin{align}
\left<\psi _{\textrm{BCS}} ^{\mathbf{b} \gamma} \right| \hat{H} \left|\psi _{\textrm{BCS}}^{\mathbf{b} \gamma}\right> &= \sum_{\mathbf{k}\neq \mathbf{b}} \epsilon _{\mathbf{k}} v_\mathbf{k}^2 + \epsilon_\mathbf{b}+ \nonumber	\\ 
	& + \sum_{\mathbf{k}{\mathbf{k'}}\neq \mathbf{b}} \left<\mathbf{k}\right|V\left|\mathbf{k'}\right>  u_\mathbf{k}v_\mathbf{k}u_{\mathbf{k'}} v_{\mathbf{k'}} ~ ,	\label{eq:bEnergyBCS}
\end{align}
\noindent where the only direct contribution of the blocked momentum state $\mathbf{b}$ to the energy is through its single particle energy. It should be noted, however, that this is not the only way that the energy depends on the blocked momentum state $\mathbf{b}$ since the distributions $v_{\mathbf{k}}$ and $u_{\mathbf{k}}$ are defined through Eqs.~(\ref{eq:bGap1BCSSwave})~{\&}~(\ref{eq:bGap2BCSSwave}) and are, therefore, dependent on $\mathbf{b}$. Solving Eqs.~(\ref{eq:bGap1BCSSwave})~{\&}~(\ref{eq:bGap2BCSSwave}) one can obtain a quasi-particle excitation energy in a way identical to Eq.~(\ref{eq:ExcEnergydefinition}). The minimum of that excitation energy will be equal to the pairing gap for the same reasons as its even-particle-number counterpart, as can be seen in Fig.~\ref{fig:N_D_MF}.

At this point, a discussion about the FSE of the pairing gap is in order. One can see very pronounced FSE in Fig.~\ref{fig:N_D_MF} for the even-particle-number and the odd-particle-number cases alike. As will be seen later, this is generally not the case for all quantities of the system. The exaggerated nature of these FSE can be understood as an interplay between the definition of the pairing gap and the quantization of the momenta: the pairing gap is defined as the minimum of the quasi-paticle excitation energy, as per  Eq.~(\ref{eq:D_MF}), and with the k-magnitudes being quantized, the position of this ``available" minimum does not necessarily fall on the true minimum that the quasi-particle excitation curve would have if it were a smooth function instead of the discretized version that we see in Fig.~\ref{fig:Exc}. Therefore, a small change in $\left<N\right>$ can shift this true minimum further from the position of the lowest point of the discretized curve making a higher or lower k-magnitude the new position of the minimum, i.e., the new lowest point of the curve. One can see clear examples of this line of thought in Fig.~\ref{fig:Exc} where it is apparent that the curves, were they smooth and continuous, would yield a minimum different than the lowest point currently circled in red.
 
As will be discussed later, an odd-particle-number system can be viewed as an one-quasi-particle excitation state of its even-particle-number vacuum. To describe the lowest of these excitations one can use the state in Eq.~(\ref{eq:bBCS}) where the momentum state $\mathbf{b}$ is chosen such that the energy of that state, namely Eq.~(\ref{eq:bEnergyBCS}), is minimum. This minimization requires a survey over the possible candidates for the state $\mathbf{b}$ where one has to solve Eqs.~(\ref{eq:bGap1BCSSwave})~{\&}~(\ref{eq:bGap2BCSSwave}) for every new $\mathbf{b}$ considered. This combined with the inherent non-linearities of Eqs.~(\ref{eq:bGap1BCSSwave})~{\&}~(\ref{eq:bGap2BCSSwave}) makes the description of such a system computationally expensive. 
\begin{figure}[htp]
\begin{center}
\includegraphics[width=1.0\columnwidth]{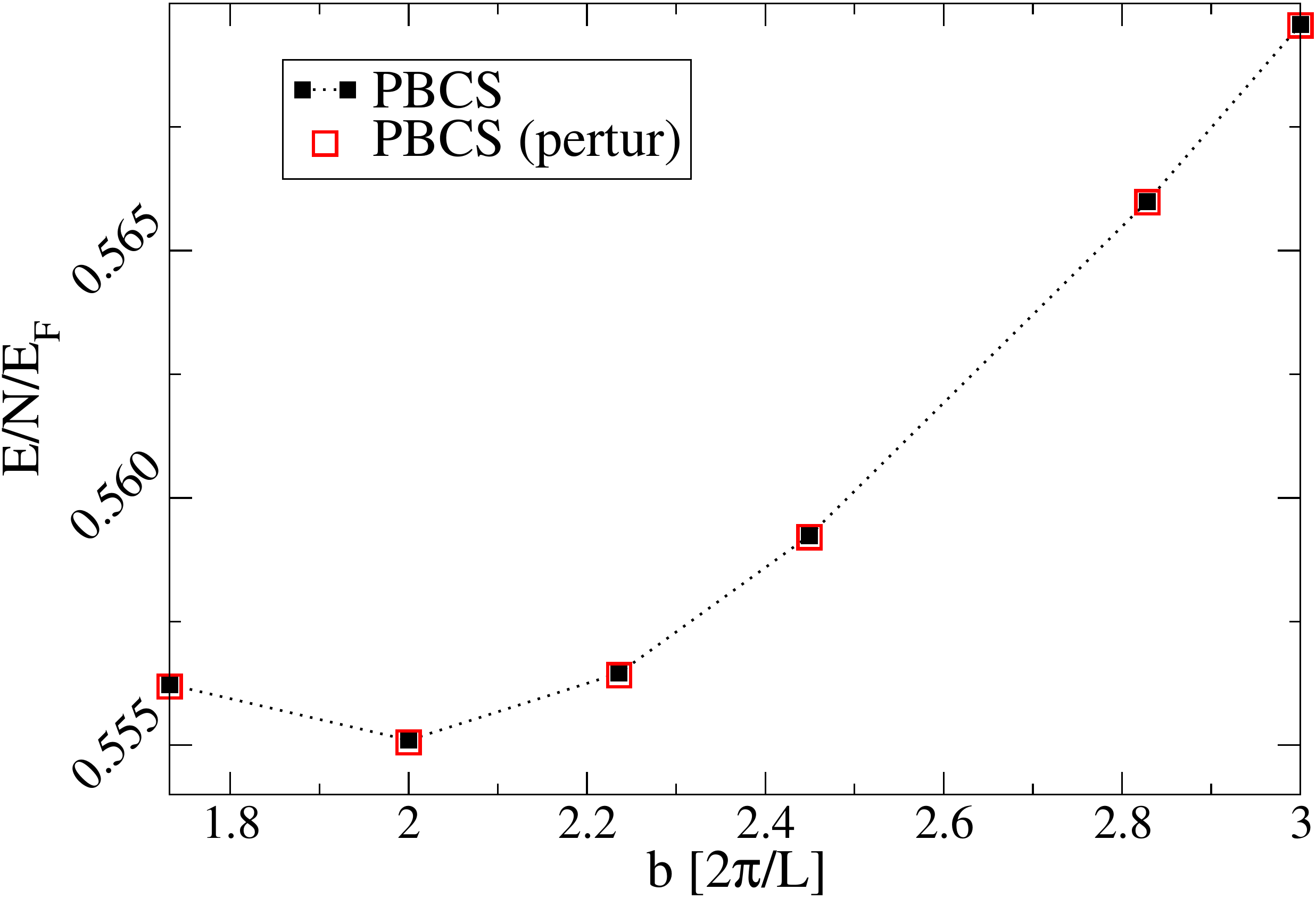}
\caption{The energy per particle, divided by the Fermi energy, of an odd-particle-number system ($\left<N\right>=67$). The energies are calculated as an optimized quasi-particle excitation of an even system and as a quasi-particle excitation of an already optimized, fully paired, wave function with $\left<N\right>=67$, at $k_{\textrm{F}}a=-10$.\label{fig:Structure}}
\end{center}
\end{figure}
An alternative to obtaining the distributions $v_{\mathbf{k}}$ and $u_{\mathbf{k}}$ by solving Eqs.~(\ref{eq:bGap1BCSSwave})~{\&}~(\ref{eq:bGap2BCSSwave}), is a perturbative approach to blocking: one can use the distributions $v_\mathbf{k}$ and $u_\mathbf{k}$ that solve Eqs.~(\ref{eq:Gap1BCSSwave})~{\&}~(\ref{eq:Gap2BCSSwave}) setting $N_0$ equal to the particle-number of the odd-particle-number state~\cite{Duguet:1:2001}, namely $N_0 = N_{\textrm{odd}}$. The error in the  distributions $v_\mathbf{k}$ and $u_{\mathbf{k}}$ resulting from a perturbative description is inversely proportional to the number of pairs $N_0/2$; such an approach is less computationally expensive since the gap equations need to be solved only once~\cite{NuclMB:Book}. Motivated by this we can use a perturbative scheme to identify the structure of the excitations of an even-particle-number system like the one in Fig.~{\ref{fig:Structure}} and compare it with the one from a non-perturbative approach. These are calculations of the energy of the state in Eq.~{(\ref{eq:bBCS})} for different blocked momenta $\mathbf{b}$. We see that the structure of the excitations remains unchanged in the sense that the energy curves produced by each approach yield a minimum at the same momentum state $\mathbf{k}$. Similar calculations for a variety of particle-numbers show the same behavior. This motivates the use of the revised perturbative scheme to locate the momentum $\mathbf{b}$ that minimizes the energy in Eq.~(\ref{eq:bEnergyBCS}) and the use of this blocked state to solve Eqs.~(\ref{eq:bGap1BCSSwave})~{\&}~(\ref{eq:bGap2BCSSwave}).

As in the even-particle-number systems, the energy of odd-particle-number systems is given by the ground state expectation value of the Hamiltonian, namely Eq.~(\ref{eq:bEnergyBCS}). Isolating the $S$-wave of the pairing interaction in this equation we arrive at the following equation for odd-particle-number systems:
\begin{align}
E^{\textrm{BCS}}_{\textrm{odd}} & (b;N) = \sum_{k\neq b} M(k)  \epsilon _{k} 2v_k^2 + \epsilon _b + 	\nonumber \\
	& + \frac{4\pi}{L^3}\sum_{k k' \neq b} M(k) M(k') V_0(k,k') u_k v_k u_{k'} v_{k'} ~ ,	\label{eq:bSwaveEnergyBCS}
\end{align}
%\begin{align}
%E^{\textrm{BCS}}_{\textrm{odd}} & (b;N) = \sum_{k} \left(M(k)  - \delta _{kb}\right)  \epsilon _{k} 2v_k^2 + \epsilon _b + 	\nonumber \\
%	& + \frac{4\pi}{L^3}\sum_{k k} \left(M(k)  - \delta _{kb}\right) \left(M(k')  - \delta _{k'b}\right)~\times \nonumber \\ 
%	&\times ~V_0(k,k') u_k v_k u_{k'} v_{k'} ~ ,	\label{eq:bSwaveEnergyBCS}
%\end{align}
\begin{figure}[htp]
\begin{center}
\includegraphics[width=1.0\columnwidth,clip=]{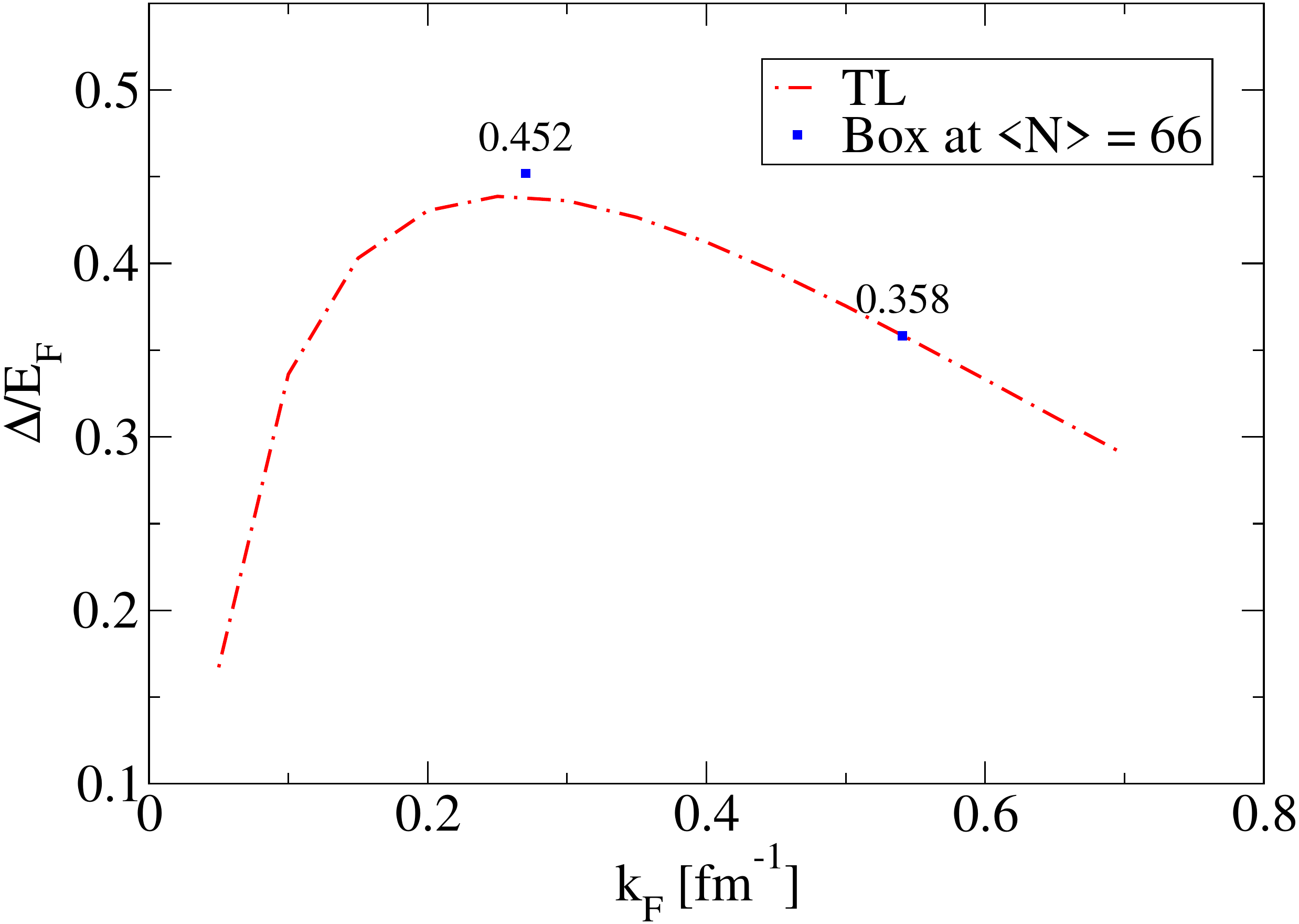}
\caption{The pairing gap at the TL in the units of the total energy as a function of the Fermi momentum $k_{\textrm{F}}$. Also graphed are the pairing gaps for $\left< N \right> =66$ at $k_\textrm{F}a=-10$ and $-5$.	\label{fig:DEf_kf}}
\end{center}
\end{figure}
\noindent where M(k) is again the population function and the quantities $v_\mathbf{k}^2$ and $v_\mathbf{k} u_\mathbf{k}v_\mathbf{k'} u_\mathbf{k'}$ have been replaced by their angle-averaged counterparts. We have also used Eq.~(\ref{eq:D2.1}). As before, the energy of the blocked state has an explicit dependence on the blocked momentum state $b$, through $\epsilon _b$, as well as an implicit one, through the distributions $u_k$ and $v_k$. In what follows, the only odd-particle-number systems that will be considered will be the ones that constitute the one quasi-particle excitation of their corresponding even-particle-number \mbox{vacuum}. Note that this distinction does not survive at the TL where $N\to \infty$ and the very distinction between even and odd particle-numbers loses its meaning. However, when discussing systems far from the TL the distinction between an even fully-paired system and its first excitation with one unpaired particle is vital in probing pairing correlations (see Sec.~\ref{sec:OES}).

\subsection{The system at the Thermodynamic Limit}

Intensive quantities for a finite system may change as the system approaches its TL, reaching their TL values as the system becomes a better approximation of the infinite one. These discrepancies of the calculated quantities from their TL value are called the finite-size effects (FSE). Typically, they are recognized as the result of a discretized $k$-space which in turn comes from the boundary conditions of a finite system and their study plays an important role in the description of infinite systems. To identify such a trend in a quantity, one needs to know the TL of the system at hand. For supefluid systems in the BCS framework this can be done straightforwardly by taking the limit of $L\to\infty$ in Eqs.~{(\ref{eq:Gap1BCSSwave})~{\&}~(\ref{eq:Gap2BCSSwave})}. That leads to the TL version of the BCS gap equations:
\begin{align}
\Delta(k) &= -\frac{1}{\pi} \int_{0}^\infty dk' (k')^2 V_0(k,k') \frac{\Delta (k')}{E(k')} ~, \label{eq:TL_SwaveGap1BCS} \\
\frac{\mv{\hat{N}}}{L^3} = n &= \frac{1}{2\pi ^2} \int_{0}^\infty dk k^2 \left(1-\frac{\xi (k)}{E(k)}\right) ~ . \label{eq:TL_SwaveGap2BCS} 
\end{align}
These equations, just like their discrete counterparts, are a set of non-linear coupled equations that have to be solved self-consistently. In Sec.~\ref{sec:Solution} we describe the way one can solve Eqs.~{(\ref{eq:Gap1BCSSwave})~{\&}~(\ref{eq:Gap2BCSSwave})} and the same method applies to Eqs.~{(\ref{eq:TL_SwaveGap1BCS})}~{\&}~(\ref{eq:TL_SwaveGap2BCS})}. Solving these one obtains the pairing gap at the TL, which can be seen in Fig.~\ref{fig:DEf_kf}, where we also plot the pairing gaps for $\left< N \right> =66$ at $k_\textrm{F}a=-10$ and $-5$. This result is consistent with the choice usually made in QMC NM calculations~\cite{Gezerlis:2010}: a periodic box of $66$ particles happens to provide a good approximation of the infinite system and can be therefore used to extrapolate to the TL.
\pagebreak[4]

Finally, an expression for the energy density at the TL can be derived by taking the limit of $L\to \infty$ in Eq.~(\ref{eq:EnergyBCS}) yielding:

\begin{align}
    \frac{E_{\textrm{TL}}}{V} =& \frac{1}{2\pi ^2}\int _0^{\infty} dk k^2 2v(k)^2 \epsilon (k) + \\ 
        &+ \frac{1}{\pi ^3} \int _0^{\infty} dk dk' k^2 {k'}^2 V_0(k,k')u(k)v(k)u(k')v(k') \label{eq:EnergyBCS_TL}~,
\end{align}

\noindent which is connected to the energy per particle as:
\begin{align}
    \frac{E_{\textrm{TL}}}{V} = n \frac{E_{\textrm{TL}}}{N}~.
\end{align}

\section{The Interaction}
\label{sec:Interaction}
At the densities considered here the NN interaction is dominated by the $^1S_0$ channel. This interaction is attractive enough to almost create a bound system (dineutron). The NN scattering length and effective range are $a\approx -18.5~ \textrm{fm}$ and $r_{\textrm{e}} \approx 2.7 ~ \textrm{fm}$, respectively~\cite{Howell:1999}. At low energies those two quantities capture the physics of the system. In other words, all potentials that can be tuned to reproduce the scattering length and effective range of NM will produce indistinguishable results at low energies regardless of the details of their functional forms (shape independence). Moreover, these parameters correspond to the free-space NN interaction. While for low density studies, such as this one, the in-medium effects can be neglected, moving to higher densities the effects of the medium have to be dealt with \cite{Ramanan:2018, Ramanan:2019} We choose to model the NN interaction with the modified P{\"o}schl-Teller potential~\cite{Gezerlis:2008}. This is:
\begin{align}
V(r) = - \frac{\hbar}{m_n}\frac{\lambda (\lambda - 1)\beta ^2}{\cosh^2{(\beta r})} ~ ,	\label{eq:PotentialPT}
\end{align}
where the parameters $\lambda$ and $\beta$ are tuned to reproduce the $^1S_0$ scattering length and effective range. In the $^1S_0$ channel the potential is:
\begin{align}
V_0 (k,k') &= \int _0 ^\infty dr r^2 j_0(kr)V(r)j_0(k'r) =  \\ 
	&=\frac{A\pi}{4\beta kk'}\left( \frac{k-k'}{\sinh{\frac{(k-k')\pi}{2\beta}}} - \frac{k+k'}{\sinh\frac{(k+k')\pi}{2\beta}} \right) \\
	 &\textrm{for} ~ k\neq k' ,\,\, k,k'\neq 0 ~, \nonumber \\ 
	&= \frac{A}{2\beta^2 k^2}\left(\beta - k\frac{\pi}{\sinh\frac{k\pi}{\beta}}\right) \\
	&\textrm{for} ~ k= k'\neq 0 ~,  \nonumber \\ 
	&= \frac{A}{2\beta^2 k^2}\left[\frac{1}{\sinh{\frac{\pi k}{2\beta}}}\left(\frac{\pi k}{2\beta} \coth{\frac{\pi k}{2\beta}} - 1\right)\right] \\
	&\textrm{for} ~ k\neq k', \,\, k' = 0 ~, \displaybreak[4] \nonumber \\
	&= \frac{A\pi ^2}{12\beta ^3} \\
	&\textrm{for} ~ k=k'=0~,  \nonumber
\end{align}
where we have defined:
\begin{equation}
    A=-\frac{\hbar}{m} \lambda (\lambda -1)\beta ^2 ~.   
\end{equation}
By choosing this form for our interaction we neglect the repulsive core of the NN interaction at short distances since the modified P{\"o}schl - Teller potential is a purely attractive potential. This repulsive core is probed at densities higher than the ones discussed here and, therefore, for our range of densities a potential with a repulsive core would produce the same results as the modified P{\"o}schl - Teller. This is consistent with the shape independence mentioned above.

\section{Solution of the BCS Gap Equations}
\label{sec:Solution}
We want to investigate the effects of pairing in the NM found in the inner crust of NSs. As noted above, our interest into the properties of the bulk medium along with the translational symmetry of the infinite medium suggests the use of a cubic box under PBC. These boundary conditions lead to the quantization of momenta:

\begin{align}
\mf{k} = \frac{2\pi}{L} \mf{n}, \quad \textrm{with}~n_x,n_y,n_z = 0,\pm 1,\dots ~,
\end{align}
which in turn leads to the single-particle energies:
\begin{align}
\epsilon _\mathbf{k} &= \frac{\hbar ^2}{2m} \left|\mathbf{k}\right| ^2 = \frac{2\pi ^2\hbar ^2}{m L^2} \left|\mathbf{n}\right| ^2~, \quad n_i = 0,\pm 1,\dots 	~. \label{PBC}
\end{align}
where the length of the box $L$ is determined so that the particle-number of the system yields the desired density.

The lack of a repulsive core (see Sec.~\ref{sec:Interaction}) in the potential permits the use of a iterative scheme for the solution of Eqs.~{(\ref{eq:Gap1BCSSwave})~{\&}~(\ref{eq:Gap2BCSSwave})}. That is, for a given value of $\mu$ we can solve Eq.~(\ref{eq:Gap1BCSSwave}) iteratively by assuming a gap distribution on the RHS and getting an updated one on the LHS. By substituting the updated gap distribution on the RHS again we get yet a new one on the LHS and so on. The iterative procedure stops when the gap distribution assumed on the RHS is equal to the updated one on the LHS. This gap distribution is the solution of Eq.~(\ref{eq:Gap1BCSSwave}) given a chemical potential $\mu$. Inserting this solution along with the given $\mu$ into Eq.~(\ref{eq:Gap2BCSSwave}) we get the average particle number that corresponds to that value of $\mu$. Using this iterative scheme we are, essentially, calculating the average particle number $\left<N\right>$ as a function of $\mu$, i.e., $\left<N\right>(\mu)$. Finally, we reduce the problem to finding the root of the equation $\left<N\right>(\mu) - N_0 = 0$ where $N_0$ is the number of particles that corresponds to the desired density in the cubic box. The TL equations are solved using the same procedure. That is, one can calculate the density as a function of $\mu$ by first solving Eq.~(\ref{eq:TL_SwaveGap1BCS}) iteratively and using its solution in Eq.~(\ref{eq:TL_SwaveGap2BCS}). More sophisticated methods have been developed for general potentials~\cite{Khodel:2001,  Ramanan:2007}. 

\section{The Particle Number Projection}
\label{sec:Projections}

As discussed above, the BCS ground state does not conserve the particle-number. One can restore the particle-number conservation by projecting out of the state in Eq.~(\ref{eq:groundstateBCS}) the component that respects this symmetry.  In the literature this is called the Projection After Variation (PAV) method or the Projected BCS (PBCS) theory. In PBCS one starts by building a self-consistent wave function in the BCS framework, as described in Sec.~\ref{sec:BCS}. A projection operator is then applied on that BCS ground state to project out the particle-conserving component of the wave function that corresponds to the right particle number, $N_0$. Earlier, following the original BCS formulation, we chose to respect this symmetry ``on average" by introducing the chemical potential in Eq.~(\ref{eq:FreeEnergy}). For systems with large particle-numbers the non-exact conservation of particles is not important since the fluctuations around the average particle number $\left<N\right>$ are of the order $1/\sqrt{N}$. However for QMC calculations where one works with finite systems of up to a hundred particles this non-conservation has to be dealt with (for a review on symmetry restoration see Ref.~\cite{Sheikh:2019}).

Using a projection operator one can project out the particle-number-conserving component of the BCS ground state~\cite{Dietrich:1964}:
\begin{align}
\left|\psi _N\right> &=\frac{1}{C}\mathcal{\hat{P}}_N \left|\psi _{\textrm{BCS}}\right> = \nonumber \\
	&= \frac{1}{C}\oint \frac{dz}{2\pi i} z^{-\frac{N}{2}-1} \prod_{\mathbf{k}} \left(u_{\mathbf{k}} + z v_{\mathbf{k}} \hat{p}_{\mathbf{k}}^{\dagger} \right) \left|0\right> = \\
	&= \frac{1}{C}\int_{0}^{2\pi} \frac{d\phi}{2\pi} e^{-i\frac{N}{2}\phi} \prod_{\mathbf{k}} \left(u_{\mathbf{k}} + e^{i\phi} v_{\mathbf{k}} \hat{p}_{\mathbf{k}}^{\dagger}\right) \left|0\right>	~ ,	\label{eq:PBCSstate}
\end{align}
\noindent where $N/2$ is the number of pairs and $C$ a normalization constant defined by requiring:
\begin{align}
\left<\psi _N|\psi _N\right> = 1~.
\end{align}

\noindent This treatment is equivalent to expressing the BCS ground state as a linear combination of eigenstates of the number operator:
\begin{align}
\left|\psi _{\textrm{BCS}}\right> &= \sum _{N}\lambda _N \left|\psi _N\right>  ~,	\label{eq:expansion}
\end{align}
and picking the one that corresponds to the right $N$ value, i.e., $N=N_0$. 
\pagebreak[4]
Using that as the ground state of the system for $N$ particles the expression for the energy becomes:
%\pagebreak[4]
\begin{align}
E^{\textrm{PBCS}}_{\textrm{even}}(N) &= \frac{\left<\psi _N\right| \hat{H} \left|\psi _N\right>}{\left<\psi _N| \psi _N\right>} = \nonumber \\
 	&=\sum_{\mathbf{k}}   \epsilon _{\mathbf{k}}2 v_{\mathbf{k}}^2 \frac{R_1^1(\mathbf{k})}{R_0^0} + \nonumber \\
 	& + \sum _{\mathbf{k}\mathbf{k'}} V_{\mathbf{k}\mathbf{l}} u_\mathbf{k} u_\mathbf{k'} v_\mathbf{k} v_\mathbf{k'} \frac{R_1^2({\mathbf{k}\mathbf{k'}})}{R_0^0} ~ ,	\label{eq:EnergyPBCS} 
\end{align}
where the quantities $R_n^m(\mathbf{k}_1\dots \mathbf{k}_m)$ are defined as the residues of contour integrals in the complex plane:
\begin{align}
 & R_n^m(\mathbf{k}_1 \mathbf{k}_2 \dots \mathbf{k}_m) = \nonumber\\
	&= \frac{1}{2\pi i} \oint dz z^{-(\frac{N}{2}-n)-1} \prod _{\mathbf{k} \neq \mathbf{k}_1, \mathbf{k} _2, \dots \mathbf{k} _m} \left( u_{\mathbf{k}}^2 + z v_{\mathbf{k}}^2\right) = 	\nonumber	 \\ 
	&= \int _0^{2\pi} \frac{d\phi}{2\pi} e^{-i(\frac{N}{2}-n)\phi} \prod _{\mathbf{k} \neq \mathbf{k}_1, \mathbf{k} _2, \dots \mathbf{k} _m} \left( u_{\mathbf{k}}^2 + e^{i\phi} v_{\mathbf{k}}^2\right)	~ ,	
	\label{eq:ResInt}
\end{align}
\noindent with $N/2$ the number of pairs described by the state in Eq.~(\ref{eq:groundstateBCS}).

From the blocked state in Eq.~(\ref{eq:bBCS}) odd-particle-number eigenstates of the number operator can be projected that describe a system with $N+1$ particles:
\begin{alignat}{2}
\left| \psi _{N+1}^{\mathbf{b} \gamma} \right> &= \frac{1}{C^{(\mathbf{b})}}\mathcal{\hat{P}}_N && \left| \psi _{\textrm{BCS}}^{\mathbf{b} \gamma}\right>  \nonumber \\
	&= \frac{1}{C^{(\mathbf{b})}}\hat{c}_{\mathbf{b}\gamma}^\dagger && \oint \frac{dz}{2\pi i} z^{-\frac{N}{2}-1} \times \nonumber	\\ 
		&	&& \times \prod_{\mathbf{k}\neq \mathbf{b}} \left(u_{\mathbf{k}} + z v_{\mathbf{k}} \hat{p}_{\mathbf{k}}^{\dagger} \right) \left|0\right>   \\ 
 	&=\frac{1}{C^{(\mathbf{b})}}\hat{c}_{\mathbf{b}\gamma}^\dagger && \int_{0}^{2\pi} \frac{d\phi}{2\pi} e^{-i\frac{N}{2}\phi} \times \nonumber \\
 	&	&& \times \prod_{\mathbf{k} \neq \mathbf{b}} \left(u_{\mathbf{k}} + e^{i\phi} v_{\mathbf{k}} \hat{p}_{\mathbf{k}}^{\dagger}\right) \left|0\right>	~ ,	\label{eq:bPBCSstate}
\end{alignat}

\noindent where $N/2$ is the number of pairs described by the state in Eq.~(\ref{eq:bBCS}) and $C^{(\mathbf{b})}$ a normalization constant defined by requiring
\begin{align}
\left<\psi _{N+1}^{\mathbf{b} \gamma}| \psi _{N+1}^{\mathbf{b} \gamma}\right> = 1~.
\end{align}

\noindent The state in Eq.~(\ref{eq:bBCS}) leads to an energy
\begin{align}
E^{\textrm{PBCS}}_{\textrm{odd}}&(\mathbf{b};N+1) = \frac{\left<\psi _{N+1}^{\mathbf{b} \gamma}\right|\hat{H}\left|\psi _{N+1}^{\mathbf{b} \gamma}\right>}{\left<\psi _{N+1}^{\mathbf{b} \gamma}|\psi _{N+1}^{\mathbf{b} \gamma}\right>} = \nonumber\\
	&= \sum_{\mathbf{k} \neq \mathbf{b}} \epsilon _{\mathbf{k}}2 v_{\mathbf{k}}^2 \frac{R_1^2(\mathbf{b}\mathbf{k})}{R_0^1(\mathbf{b})} + \epsilon _{\mathbf{b}}  + \nonumber \\
	& + \sum _{\mathbf{k}\mathbf{k'}\neq \mathbf{b}} V_{\mathbf{k}\mathbf{k'}} u_\mathbf{k} u_\mathbf{k'} v_\mathbf{k} v_\mathbf{k'} \frac{R_1^3({\mathbf{b}\mathbf{k}\mathbf{k'}})}{R_0^1(\mathbf{b})}	~ . \label{eq:bEnergyPBCS} 
\end{align}
The residuum integrals can be calculated numerically using Eq.~(\ref{eq:ResInt}). 
\pagebreak[4]
These two prescriptions for the calculation of the energies of even-particle-number and odd-particle-number finite superfluid systems, namely Eqs.~(\ref{eq:EnergyPBCS})~{\&}~(\ref{eq:bEnergyPBCS}), constitute a beyond-mean-field treatment since the expansion in Eq.~(\ref{eq:expansion}) introduces correlations beyond-mean-field~\cite{Sheikh:2019}. Finally, as with the BCS treatment, the $S$-wave component of the interaction has to be isolated from these energy expressions leading to the following expression for even-particle-number systems:
%\pagebreak[4]
\begin{align}
&E^{\textrm{PBCS}}_{\textrm{even}}(N) = \sum_{k} M(k) \epsilon _{k} 2v_k^2 \frac{R_1^1(k)}{R_0^0}+ 	\nonumber \\
	&+ \frac{4\pi}{L^3} \sum_{k k'}  M(k)M(k') V_0(k,k')~\times \nonumber \\
	& \times ~u_kv_k u_{k'} v_{k'} \frac{R_1^2(kl)}{R_0^0}	~ ,	 \label{eq:SwaveEnergyPBCS}
\end{align}
\noindent and for odd-particle-number-systems:
\begin{align}
&E^{\textrm{PBCS}}_{\textrm{odd}}(b;N) = \sum_{k\neq b} M(k) \epsilon _{k} 2v_k^2 \frac{R_1^2(bk)}{R_0^1(b)} +  	\nonumber \\
	& + \epsilon _b + \frac{4\pi}{L^3} \sum_{k k'\neq b} M(k)M(k')  V_0(k,k') ~\times \nonumber \\
	&\times~ u_k v_k u_{k'} v_{k'} \frac{R_1^3(bkl)}{R_0^1(b)}	~ .\label{eq:bSwaveEnergyPBCS}
\end{align}

\begin{figure}
\begin{center}
\includegraphics[width=1.0\columnwidth,clip=]{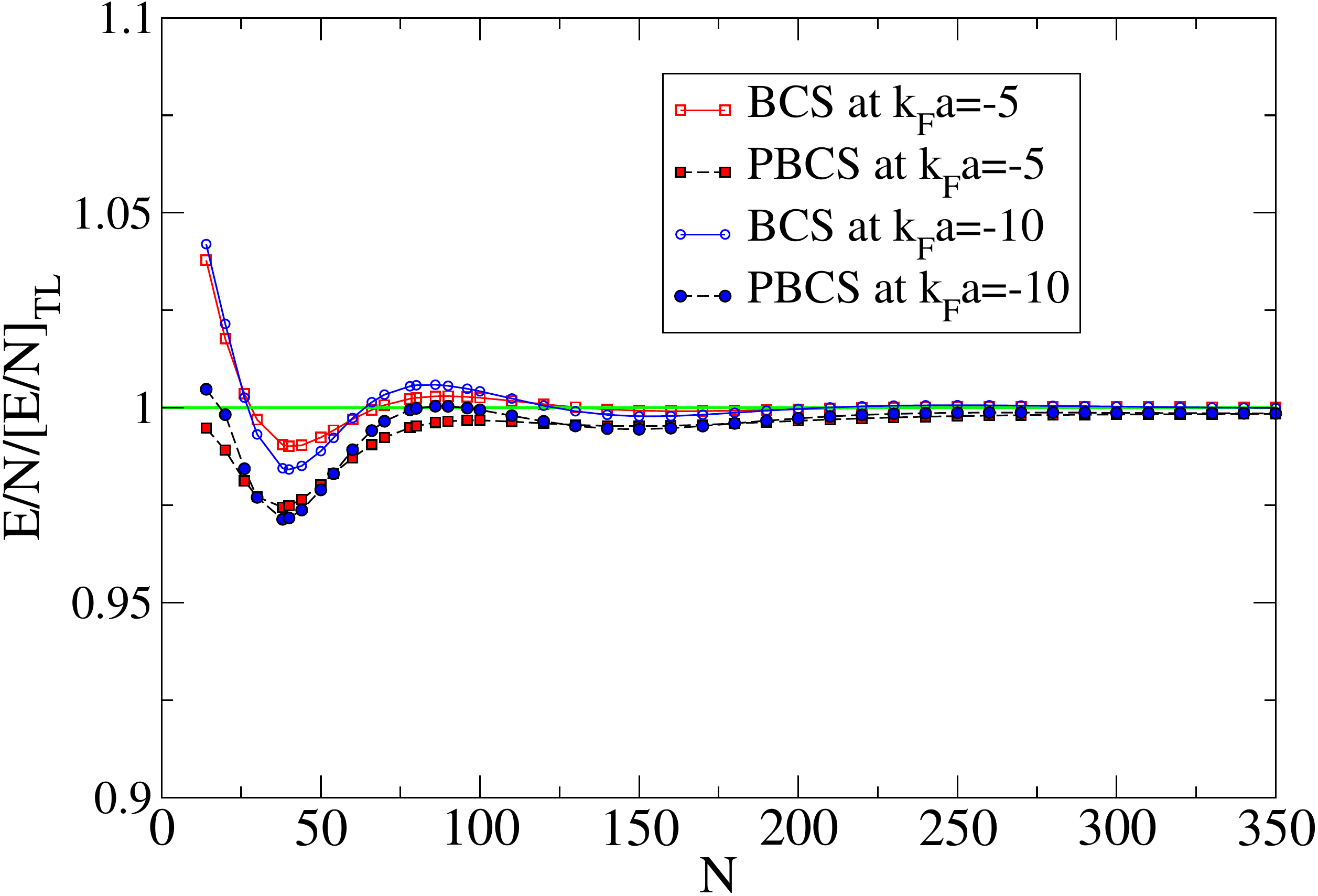}
\caption{The energy in BCS and PBCS as a function of the particle number $N$ for $k_F a = - 10$ and $- 5$.	\label{fig:N_E}}
\end{center}
\end{figure}
\noindent Using Eqs.~{(\ref{eq:SwaveEnergyPBCS})~{\&}~(\ref{eq:bSwaveEnergyPBCS})} one can calculate the energy of a system with a fixed number of particles $N$. A calculation of that energy can be seen in Fig.~\ref{fig:N_E} where we plot the energy of an even-particle-number system for $k_{\textrm{F}}a=-10$ and $-5$ in BCS and PBCS, that is, using Eqs.~{(\ref{eq:SwaveEnergyBCS})~{\&}~(\ref{eq:SwaveEnergyPBCS})} respectively. It is worth noting that the FSE for the energy are smaller than for the pairing gap as can be seen in Fig.~{\ref{fig:N_D_MF}}.

It should also be noted that the projection method described above starts by building a self-consistent BCS wave function, namely the state in Eq.~(\ref{eq:groundstateBCS}) (or the state in Eq.~(\ref{eq:bBCS}) for odd-particle-number systems). That BCS ground state, however, is built to minimize the free energy in Eq.~(\ref{eq:FreeEnergy}) and that does not guarantee that the resulting projected state in Eq.~(\ref{eq:PBCSstate}) (or Eq.~(\ref{eq:bPBCSstate}) for odd-particle-number systems) will also be self-consistent. That is, if one were to determine the distributions $v_\mathbf{k}$ and $u_\mathbf{k}$ so that they minimize the energy in Eq.~(\ref{eq:EnergyPBCS}) (or Eq.~(\ref{eq:bEnergyPBCS}) for odd-particle-number systems) they would, in principle, find different distributions than the ones that come from the BCS treatment described in Sec.~\ref{sec:BCS}. This latter approach of minimizing the energy of the projected state is called the Variation After Projection Method (VAP) and it has been used to restore the particle-number symmetry for $nn$ and $pp$ pairing in nuclei~\cite{Mang:1965} as well as the particle-number, spin and isospin symmetries in $np$ pairing~\cite{Romero:2019}. It can be shown that for strong pairing the VAP and BCS descriptions are equivalent~\cite{Dietrich:1964}. Furthermore, the PBCS approach gets closer to the VAP one as the pairing correlations increase. In this work we are interested in the pairing correlations in NM which exhibits some of the strongest pairing effects in nature. Therefore we extend our BCS results to finite systems using the PBCS theory. 

\section{The Pairing Gap and the Odd-Even Staggering}
\label{sec:OES}
The key signature of pairing correlations is the occurrence of the so-called pairing gap. The pairing gap is manifested in two different observables. First, a gap is observed in the quasi-particle excitation spectrum of the BCS ground state (see Fig.~\ref{fig:Exc}) and, second, there is an energy shift between the energy curves of even-particle-number and odd-particle-number systems. Both of those features can be exploited to calculate the pairing gap.

The first effect mentioned above lets us identify the minimum of the quasi-particle excitation energy as the pairing gap (cf. Eq.~{(\ref{eq:D_MF})}). The value of this minimum, even though not explicitly stated in Eq.~(\ref{eq:D_MF}), is also a function of the average particle number as can be seen in Fig.~\ref{fig:Exc}.

The second effect mentioned above can be also exploited to calculate the pairing gap by point-wise interpolating the two curves using finite-difference formulae and calculating the shift for each particle number $N$. This treatment, inspired by the odd-even mass staggering in nuclei, leads to formulae containing only differences of the energy of systems with different particle-numbers. Additionally, increasing the level of sophistication with which the pairing has been dealt in each of the energy calculations that go into the odd-even staggering formulae, one can get beyond-mean-field contributions of increasing accuracy. Furthermore, different interpolation schemes give rise to OES formulae of different orders. Beyond-mean-field correlations, resulting from more than mere particle-projection have been studied in nuclei\cite{Satula:1998, Heenen:2001, Neergard:2019}. However, this study, as mentioned above, has a more limited scope which is to provide guidance for \textit{ab initio} approaches that will presumably capture such beyond-mean-field effects accurately. Therefore we investigate the phenomenological value of the odd-even staggering and not the correlations induced by a particle-number projection. In studies of superfluid systems, for a given choice of Hamiltonian and boundary conditions, the odd-even staggering is a quantity that can be tackled in a variety of theoretical approaches and so it can help compare their results without any dependence on the details of the formalism. Thus, the connection of odd-even staggering to phenomenological approaches, like the BCS treatment, is important and should be well-defined. 

Finite-difference formulae are derived from a Taylor series expansion of the energy as a function of the particle-number~\cite{Bender:2000}:

\begin{align}
E(N) = \sum \frac{1}{n!}\frac{\partial ^n E_0}{\partial N^n} \bigg|_{N_0} \left(N-N_0\right)^n + D(N)	~ ,	\label{eq:Taylor}	
\end{align}  

\noindent where $E_0(N)$ is the energy of a fully paired BCS wave function and $D(N)$ is the gap defined as:

\begin{align}
D(N) = \branch{0}{\textrm{for even } N}{\Delta (N)}{\textrm{for odd } N }	~ . \label{eq:DN}
\end{align}

\noindent The energy $E_0 (N)$ corresponds to the energy in Eq.~(\ref{eq:EnergyBCS}) (or Eq.~(\ref{eq:EnergyPBCS}) for PBCS) where $v_\mathbf{k}$ and $u_\mathbf{k}$ come from solving the BCS gap equations, namely Eqs.~(\ref{eq:Gap1BCSSwave})~{\&}~(\ref{eq:Gap2BCSSwave}), setting $\left<N\right>$ equal to the even or odd particle-number. Note that for odd-particle-number systems, this energy does not necessarily correspond to the energy in Eq.~(\ref{eq:bEnergyBCS}) (or Eq.~(\ref{eq:bEnergyPBCS}) for PBCS) where $v_\mathbf{k}$ and $u_\mathbf{k}$ come from solving the BCS gap equations (\ref{eq:bGap1BCSSwave})~{\&}~(\ref{eq:bGap2BCSSwave}) setting $\left<N\right>$ equal to the odd particle number. Denoting it $E_{\textrm{blocked}}$, the former energy is:
\begin{align}
E_{\textrm{blocked}} = E_0(N) + \Delta ^{(b)} (N) ~ ,	  
\end{align}

\noindent where $\Delta ^{(b)}$ is a quantification of blocking to which we will refer to as the ``blocking gap". It is not the same as the pairing gap as, along with pure pairing correlations, it contains the polarization effects that arise from the breaking of the time-reversal symmetry by blocking the momentum state $\mathbf{b}$ in the wave function in Eq.~(\ref{eq:bBCS}).

The OES formulae for the gap $\Delta (N_0)$ are linear combinations of values of $E(N)$ for $N$ around $N_0$ where the contributions of $E_0(N)$ and its first $2M-1$ derivatives vanish. Their construction can be found in Appendix \ref{App:OES}. Using Eq.~{(\ref{eq:BMfinal})} for different values of $M$ we get OES formulae of different orders. For $M=1$, we get the three-point (second order) OES expression:
\begin{align}
\Delta (N_0) = - & \frac{(-1)^{N_0}}{2} \big[2E(N_0) - E(N_0+1) - E(N_0 -1)\big]	~ ,	\label{eq:3pOES}
\end{align}
while for $M=2$ we get the five-point (fourth order) OES expression
\pagebreak[4]
\begin{align}
\Delta (N_0) = &-\frac{(-1)^{N_0}}{8} \bigg[ E(N_0+2) - 4E(N_0 +1) + 6E(N_0) 	\nonumber \\
	&-4E(N_0-1) + E(N_0-2) \bigg]	~ .\label{eq:5pOES}
\end{align}
These finite-difference formulae can also be understood as an estimation of the shift between the even-$N$ and odd-$N$ curves interpolating from given points to different orders. These are the most widely used expressions in the literature, along with a hybrid five-point formula and a four-point formula~\cite{Mang:1965}.

The OES treatment aims to decouple the pure pairing correlations from the underlying mean-field treatment that is used to calculate $E(N)$. Naturally, higher order differences result in better decoupling. Additionally, extra mean-field contributions could leak into a pairing gap calculated by an OES formula when $N_0$ is not an odd number~\cite{Duguet:2:2001}.  

\begin{figure}[htp]
\begin{center}
\includegraphics[width=1.0\columnwidth,clip=]{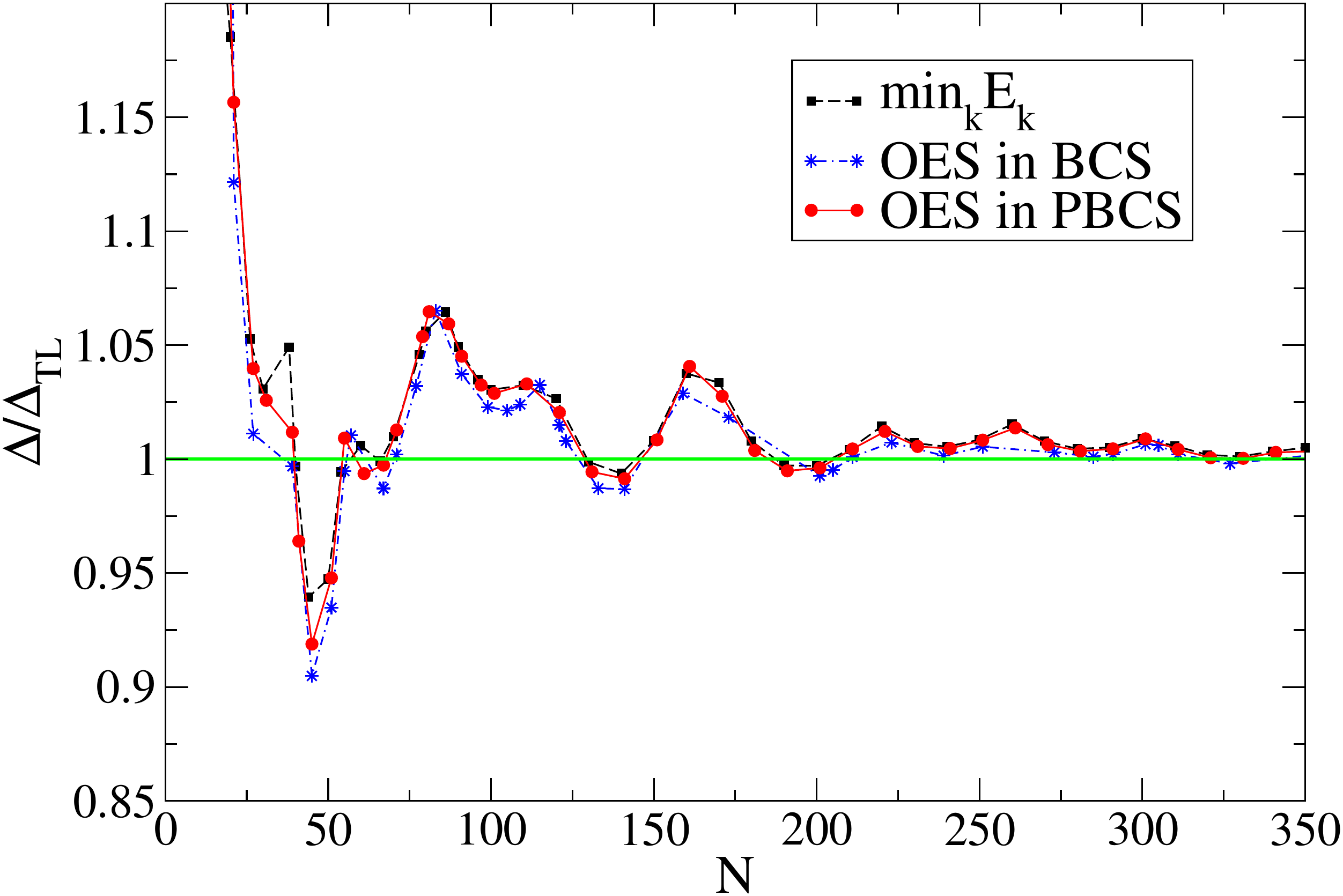}
\caption{The pairing gap at $k_F a = -10$ as a result of a mean-field and the OES treatment.\label{fig:OES}}
\end{center}
\end{figure}

We calculated the pairing gap in NM using the three-point formula in Eq.~(\ref{eq:3pOES}) for densities ranging from $k_F a = -5$ to $k_F a = -10$. Equation~(\ref{eq:3pOES}) was centered around odd particle-numbers to minimize the mean-field contributions. The odd-particle-number energies $E(N)$ correspond to the energies of an one quasi-particle excitation states of the even-particle-number vacuum of energy $E(N-1)$ and they were obtained using a mix of the revised perturbative scheme and the self-consistent method as described in Sec.~\ref{sec:BCS}. Calculations were done with energies coming from both BCS and PBCS treatments with their results being in reasonable agreement as can be seen in Fig.~\ref{fig:OES}. It should be noted that an OES treatment in the context of BCS is, by definition, ill-defined since the energy $E(N)$ in BCS refers to the average energy of an ensemble of systems with an average particle-number $\left<N\right>=N$. However, it is shown here for the sake of completeness. 

\begin{figure}
\begin{center}
\includegraphics[width=1.0\columnwidth,clip=]{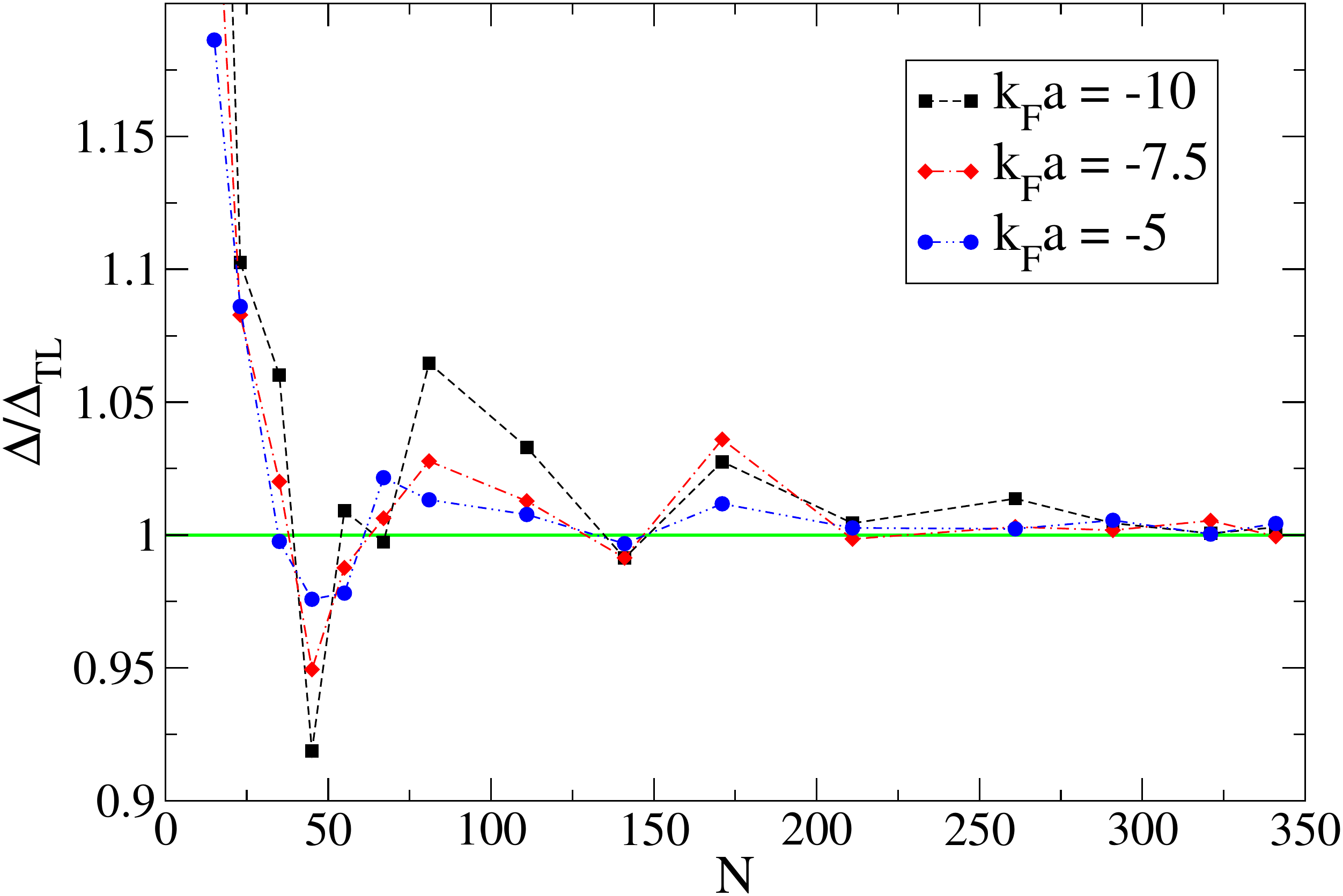}
\caption{The pairing gap at $k_F a = -10, -7.5$ and $-5$ as a result of the OES treatment.\label{fig:OES_Master}}
\end{center}
\end{figure}

Evidently, from Fig.~\ref{fig:OES}, the pairing gap resulting from the OES treatment is in agreement with the mean-field gap from Eq.~(\ref{eq:D_MF}). In more detail, we see that in the region of $N=66$ the OES treatment is equivalent to the mean-field one and both of them are good approximations of the TL. This motivates and further justifies the study of pairing in NM in systems of $N=66$ particles, a practice that has been standard in \textit{ab initio} studies of NM \cite{Gandolfi:2015, Gezerlis:2014, Hagen:2013} due to the shell closure of the free Fermi gas. This behavior is not surprising: a system with a short-range interaction at low densities ($L \gg r_\textrm{e}$) is expected to be similar to the non-interacting one, at least as far as the FSE are concerned. At larger densities, the validity of $N=66$ as a good approximation of the TL has been verified by \textit{ab initio} studies of NM \cite{Gezerlis:2014, Hagen:2013}. A comparison between the pairing gap from OES at $k_F a = -10, -7.5$, and $-5$ can be seen in Fig.~\ref{fig:OES_Master}. Higher order treatments such as the one in Eqs.~(\ref{eq:5pOES}), were also used giving results identical to that of the three-point-formula indicating that a second order approximation to the pairing is of sufficient accuracy.
\pagebreak[4]

The mean-field pairing gap from the BCS theory and the beyond-mean-field gap from the PBCS theory are expected to agree at the TL.  That is, Eqs.~(\ref{eq:D_MF})~{\&}~(\ref{eq:3pOES}) are equivalent as $N \to \infty$. That can be seen clearly in Fig.~\ref{fig:OES} where the BCS gap and the PBCS gap reach the same value as the number of particles increases. However, what is also evident from the same figure is that this agreement is present far from the TL as well.

\section{Summary \& Conclusions}
\label{sec:Summary}

In summary, we performed calculations of the pairing gap in pure NM for a range of densities relevant to the inner crust of NSs using a realistic interaction tuned to reproduce the scattering length and effective range of the bare NN interaction. The calculations were done in the BCS framework where one treats the system as a part of a grand-canonical ensemble. We also performed a symmetry restoration to recover the lost particle-number symmetry and get wave functions that describe the finite system more accurately. Finally in the context of the symmetry-restored theory (PBCS) we calculated the pairing gap using OES prescriptions. Our work shows that, far from the TL, the pairing gap as a result of a mean-field treatment matches the pairing gap calculated through OES. Moreover, OES formulae of different orders of accuracy are in good agreement with each other indicating that the NM pairing correlations in the $^1S_0$ channel can be captured by the OES of the lowest order, namely the 3-point formula. 

Our findings motivate the study of pairing in NM in systems of $N=66$ particles~\cite{Gezerlis:2010} where the two approaches agree with each other and provide a good approximation of the TL (see Fig.~\ref{fig:OES}). Given the large scattering length of NM ($a\approx-18.5~\textrm{fm}$), one can extend these results to cold Fermi atoms via the unitarity regime, to the extent that the finite effective range of NM can be neglected. Quantum Monte Carlo techniques can be used to carry out studies for both NM and cold Fermi atoms at great precision~\cite{Gezerlis:2008} as well as studies of the unitarity regime that connects them~\cite{Carlson:2003,Jensen:2020}. Given our results, one can connect such studies to other techniques in the literature.

\section*{Acknowledgements}
A.G. would like to acknowledge insightful conversations with T. Duguet. This work was supported by the Natural Sciences and Engineering Research Council (NSERC) of Canada, the Canada Foundation for Innovation (CFI), and the Early Researcher Award (ERA) program of the Ontario Ministry of Research, Innovation and Science. Computational resources were provided by SHARCNET and NERSC.

\appendix

\section{The partial wave expansion of the BCS gap equations}
\label{App:SwaveGap}
In this appendix we will present the $S$-wave expansion of functions of vectors that are in the form of sums of the potential multiplied with other functions of vectors. We will use the final expressions in our $S$-wave expansion of the gap equations where the vectors are momenta on a $3$D $\mfk$-lattice (see Eqs.~(\ref{eq:Gap1BCSSwave})~{\&}~(\ref{eq:Gap2BCSSwave})) or in a continuum $3$D $\mfk$-space (see Eqs.~(\ref{eq:TL_SwaveGap1BCS})~{\&}~(\ref{eq:TL_SwaveGap2BCS})). The momenta in the following derivation are denoted by general vectors $\vn$ in an attempt to hint that those expressions are applicable outside BCS theory as well.

\subsection{Single-sum quantities}
Let $S(\vn)$ be a quantity that depends on $\vn$ in the following way:
\begin{align}
S(\vn) = \sum _{\vn '} \bra{\vn} V \ket{\vn '} B(\vn ')	~ ,	
\end{align}
where $B( \vn )$ is an arbitrary function of $\vn$. We can separate the dependencies of the quantity $S$ from different channels of the potential and eventually express it as:
\begin{align}
S(\vn) = \sum_l S_l	(\vn)~ .
\end{align}
\noindent To that end we separate the potential's radial and angular dependence on the momenta:
\begin{align}
\bra{\vn} V \ket{\vn '} = \frac{4\pi}{L^3} \sum_{l=0}^\infty (2l+1) V_l(\nu, \nu ') P_l(\vne \cdot \vne ')	~ ,	
\end{align}
where
\begin{align}
V_l(\nu,\nu ') &= \int dr r^2 j_l(k_\nu r) V(r) j_l(k_{\nu '} r) ~,
\end{align}
and {$P_l$} are {the Legendre Polynomials} and {$j_l$}~{the spherical Bessel functions}. Using that, the quantity $S$ becomes
\begin{align}
S(\vn) &= \sum _{\vn '} \frac{4\pi}{L^3} \sum_{l=0}^\infty (2l+1) V_l(\nu, \nu ') P_l(\vne \cdot \vne ') B(\vn ') 	\nonumber	\\
	&=   \sum_{l=0}^\infty \frac{4\pi}{L^3}\sum _{\vn '} [(2l+1) V_l(\nu, \nu ') P_l(\vne \cdot \vne ')] B(\vn ')  	\nonumber	\\
	&= \sum_l S_l (\vn)	~ ,	
\end{align}
where
\begin{align*}
S_l (\vn) = \frac{4\pi}{L^3}\sum _{\vn '} [(2l+1) V_l(\nu, \nu ') P_l(\vne \cdot \vne ')] B(\vn ') \label{eq:S1} 	~ ,	
\end{align*}

\noindent and therefore:
\begin{align}
S_0(\vn) &= \frac{4\pi}{L^3}\sum _{\vn '} [V_0(\nu, \nu ') P_0(\vne \cdot \vne ')] B(\vn ')  	\nonumber 	\\
	&=\frac{4\pi}{L^3}\sum _{\vn '} V_0(\nu, \nu ') B(\vn ') ~	. 
\end{align}

\noindent Having separated $S$ into different channels we can also separate the radial and angular dependencies of $S$. To do so we can use the expansion of the Legendre polynomials in spherical harmonics:
\begin{align}
P_l(\vne \cdot \vne ') = \frac{4\pi}{2l+1} \sum_{m=-l}^l Y_{lm}(\vne) Y_{lm}^*(\vne ') ~ .	\label{eq:*}
\end{align}
With that, $S$ becomes
\begin{align}
S(\vn) &= \sum_{lm} \frac{4\pi}{L^3}\sum _{\vn '} [4\pi V_l(\nu, \nu ')Y_{lm}(\vne) Y_{lm}^*(\vne ')] B(\vn ')  \nonumber 	\\
&= \sum_{lm} \sqrt{\frac{4\pi}{2l+1}} Y_{lm}(\vne) \times \nonumber \\ 
	&\left[ \frac{4\pi}{L^3}\sum _{\vn '} (2l+1)\sqrt{\frac{4\pi}{2l+1}}  Y_{lm}^*(\vne ') V_l(\nu, \nu ') B(\vn ')\right]	~ .
\end{align}

\noindent Expanding $S$ in a Laplace series:
\begin{align}
S(\vn) = \sum _{lm} \sqrt{\frac{4\pi}{2l+1}} Y_{lm}(\vne) S_{lm} (\nu)	~ ,	
\end{align}
and using the orthogonality of the spherical harmonics we can identify $S_{lm}$ as
\begin{align}
S_{lm}(\nu) = \frac{4\pi}{L^3}\sum _{\vn '} & (2l+1)\sqrt{\frac{4\pi}{2l+1}} Y_{lm}^*(\vne ') V_l(\nu, \nu ') B(\vn ') ~ ,	\label{eq:S1*} 
\end{align}
and therefore
%\pagebreak[4]
\begin{align}
S_{00} (\nu) &= \frac{4\pi}{L^3}\sum _{\vn '} \sqrt{4\pi}  Y_{00}^*(\vne ') V_0(\nu, \nu ') B(\vn ') 	\nonumber \\
	&= \frac{4\pi}{L^3}\sum _{\vn '} V_0(\nu, \nu ') B(\vn ') ~ ,	\label{eq:S1.1star} 
\end{align}
At this point we can also identify the ways that $S$ depends on different moments of $B$. To do so we have to expand $B$ in its own Laplace series,
\begin{align}
B(\vn ') = \sum _{l'm'} \sqrt{\frac{4\pi}{2l'+1}} Y_{l'm'}(\vne ') B_{l'm'} (\nu ')	~ .
\end{align}

\noindent If we substitute that in the expression for $S$ we get
\begin{align}
S(\vn) &=\sum _{\vn '} \frac{4\pi}{L^3} \sum_{l=0}^\infty (2l+1) V_l(\nu, \nu ') \frac{4\pi}{2l+1}  \sum_{m=-l}^l Y_{lm}(\vne) \times 	\nonumber	 \\
	&\times Y_{lm}^*(\vne ')\sum _{l'm'} \sqrt{\frac{4\pi}{2l'+1}} B_{l'm'} (\nu ')Y_{l'm'}(\vne ') 	\nonumber	\\
&= \sum_{lm}\sqrt{\frac{4\pi}{2l+1}} Y_{lm}(\vne) \bigg[ \frac{4\pi}{L^3} \sum _{\vn '}\sum _{l'm'} 4\pi \sqrt{\frac{2l+1}{2l'+1}} \times 	\nonumber	\\
 	&\times Y_{lm}^*(\vne ') Y_{l'm'}(\vne ') V_l(\nu, \nu ') B_{l'm'} (\nu ') \bigg]	~ .
\end{align}
Again, using the orthogonality of the spherical harmonics we can identify $S_{lm}$ as
\begin{align}
S_{lm}(\nu) &= \frac{4\pi}{L^3} \sum _{\vn '}\sum _{l'm'} 4\pi \sqrt{\frac{2l+1}{2l'+1}}  \times 	\nonumber	 \\
	&\times  Y_{lm}^*(\vne ') Y_{l'm'}(\vne ') V_l(\nu, \nu ') B_{l'm'} (\nu ') ~ .	\label{eq:S2} 
\displaybreak[4]
\end{align}
Therefore
\begin{alignat}{2}
S_{00} =& \frac{4\pi}{L^3} \sum _{\vn '}\sum _{l'm'} &&  4\pi \sqrt{\frac{1}{2l'+1}}  \times 	\nonumber	 \\
	&	&&\times Y_{00}^*(\vne ') Y_{l'm'}(\vne ') V_0(\nu, \nu ') B_{l'm'} (\nu ')  	\nonumber\\
 =& \frac{4\pi}{L^3}\bigg\{ \sum_{\nu '} V_0 && (\nu, \nu ') B_{00} (\nu ') +  \nonumber	 \\
	& + \sum _{\vn '}{\sum _{l'm'}}^\prime && \sqrt{\frac{4\pi}{2l'+1}} Y_{l'm'}(\vne ') V_0(\nu, \nu ') B_{l'm'} (\nu ') \bigg\} ~.
\end{alignat}
where
\begin{align*}
{\sum_{l'm'}}^\prime = \sum_{l'=1}^\infty \sum_{m'=-l'}^{l'}	~ .
\end{align*}

\noindent Finally keeping only the $B_{00}$ term:
\begin{align}
S_{00} =& \frac{4\pi}{L^3} \sum_{\nu '} V_0(\nu, \nu ') B_{00} (\nu ') ~ .\label{eq:2.1} 
\end{align}

\subsection{Double-sum quantities}

\noindent Let $Q(\vn)$ be a quantity that depends on $\vn$ in the following way:
\begin{align}
Q(\vn) = \sum _{\vn ',\vn  ''} \bra{\vn '} V \ket{\vn  ''} B(\vn ',\vn  '',\vn)	~ ,	
\end{align}

\noindent where $B(\vn ',\vn  '',\vn)$ is an arbitrary function of $\vn ',\vn  '',\vn$. Carrying out a similar derivation as the one presented for the single-sum quantities we have to separate the radial and angular dependencies of the potential,
\begin{align}
\bra{\vn '} V \ket{\vn  ''} = \frac{4\pi}{L^3} \sum_{l=0}^\infty (2l+1) V_l(\nu', \nu '') P_l(\vne ' \cdot \vne '')~ .
\end{align} 
Plugging this in we get
\begin{align}
Q(\vn) &= \sum_{l=0}^\infty \frac{4\pi}{L^3}\sum _{\vn ',\vn  ''} [(2l+1) V_l(\nu ', \nu '') \times \nonumber \\
	&\times P_l(\vne ' \cdot \vne '')]B(\vn ',\vn  '',\vn)  \nonumber 	\nonumber \\
	&=\sum_l Q_l (\vn)	~ ,	
\end{align}
where
\begin{align}
Q_l(\vn) = \frac{4\pi}{L^3}\sum _{\vn ',\vn  ''} [(2l+1) & V_l(\nu ', \nu '') P_l(\vne ' \cdot \vne '')] \times \nonumber \\
	&B(\vn ',\vn  '',\vn) ~ ,	\label{eq:D1} 
\end{align}
and therefore
\begin{align}
Q_0(\vn) &= \frac{4\pi}{L^3}\sum _{\vn ',\vn  ''} [V_0(\nu ', \nu '') P_0(\vne ' \cdot \vne '')]B(\vn ',\vn  '',\vn)   \nonumber	\\
	&=\frac{4\pi}{L^3}\sum _{\vn ',\vn  ''} V_0(\nu ', \nu '')B(\vn ',\vn  '',\vn)	~ . \label{eq:D1.1} 
\end{align}

\noindent Following similar steps as before, we can separate the radial and angular $\vn$-dependences of $B$ by expanding it in a Laplace series,
\begin{align}
B(\vn ',\vn  '',\vn) = \sum _{lm} \sqrt{\frac{4\pi}{2l+1}} Y_{lm}(\vne) B_{lm}(\vn ',\vn  '',\nu)	~ ,	
\end{align}

\noindent using that in the expression for $Q$ and using the expansion of the Legendre Polynomials, namely Eq.~(\ref{eq:*}), we get
\begin{align}
Q(\vn) &= \sum _{lm} \sqrt{\frac{4\pi}{2l+1}} Y_{lm}(\vne) \bigg\{ \frac{4\pi}{L^3} \sum _{\vn ',\vn  ''} \sum_{l'm'} V_l'(\nu ', \nu '')  \times 	\nonumber	 \\
	&\times Y_{l'm'}(\vne ') Y_{l'm'}^*(\vne '') B_{lm}(\vn ',\vn  '',\nu) \bigg\} 	~ .
\end{align}

\noindent Expanding $Q$ in its own Laplace series and using the orthogonality of the spherical harmonics we identify $Q_{lm}$ as
\begin{align}
Q_{lm}(\nu) &= \frac{4\pi}{L^3} \sum _{\vn ',\vn  ''} \sum_{l'm'} V_l'(\nu ', \nu '')  \times 	\nonumber	 \\
	&\times Y_{l'm'}(\vne ') Y_{l'm'}^*(\vne '') B_{lm}(\vn ',\vn  '',\nu) 	\nonumber	\\
 &= \frac{4\pi}{L^3} \sum _{\vn ',\vn  ''} \bigg[V_0(\nu ', \nu '') +  {\sum_{l'm'}}^\prime V_l'(\nu ', \nu '')  \times 	\nonumber	 \\
	&\times Y_{l'm'}(\vne ') Y_{l'm'}^*(\vne '')\bigg] B_{lm}(\vn ',\vn  '',\nu) ~ ,	\label{eq:D2} 
\end{align}
 therefore
\begin{align}
Q_{00}(\nu) = \frac{4\pi}{L^3} \sum _{\vn ',\vn  ''} V_0(\nu ', \nu '') B_{00}(\vn ',\vn  '',\nu)	~ .	\label{eq:D2.1} 
\end{align}

\section{The residuum integrals}
\label{App:ResInt}
For convenience, in this section we will adopt the following notation:
\begin{align}
\duo{k} &= \hat{c}_{\mathbf{k} \uparrow} \hat{c}_{-\mathbf{k} \downarrow} \nonumber 	\nonumber 	\\
\duo{k}^\+ &= \hat{c}_{-\mathbf{k} \downarrow}^\+ \hat{c}_{\mathbf{k} \uparrow} ^\+	~ .
\end{align}

\noindent It can be shown that:
\begin{align}
\comm{\duod{k}}{\cts{l}} &= \ctu{k}^\+ \delta_{\mf{l},-\mfk} \delta _{\sigma \downarrow} - \mctd{k} ^\+ \delta_{\mf{l},\mfk}\delta_{\sigma \uparrow} \\
\comm{\duo{k}}{\cts{l}^\+} &=  \mctd{k} \delta_{\mf{l},\mfk}\delta_{\sigma \uparrow}-\ctu{k} \delta_{\mf{l},-\mfk} \delta _{\sigma \downarrow}  \\
\comm{\duod{k}}{\cts{l}^\+} &= \comm{\duo{k}}{\cts{l}} = 0  \\ 
\comm{\duod{k}}{\duo{l}} &= \left( \nks{k}{\uparrow} + \nks{k}{\downarrow} - 1 \right) \delta_{\mf{l},\mfk}  \\ 
\comm{\duo{k}}{\duo{l}} &= \comm{\duod{k}}{\duod{l}} = 0	~ .
\end{align}
Starting with the normalization of the even-particle-number wave function in PBCS,
\pagebreak[4]
\begin{align}
\left<\psi _N|\psi _N\right> = \frac{1}{\left|C\right|^2} &\int \frac{d\phi _1}{2\pi} \int \frac{d\phi _2}{2\pi} e^{-i\frac{N}{2}(\phi _2 -\phi _1)} \times \nonumber \\
	&\times \left<0\right|\prod_{\mathbf{l}}\left(u_{\mathbf{l}} + v_{\mathbf{l}} e^{-i\phi _1} \hat{p}_{\mathbf{l}} \right)  \times \nonumber \\
	&\times  \prod_{\mathbf{k}} \left( u_{\mathbf{k}} + v_{\mathbf{k}} e^{i\phi _2} \hat{p}_{\mathbf{k}}^{\dagger} \right) \left|0\right>  \nonumber \\
	= \frac{1}{\left| C\right|^2} \int &\frac{d\phi _1}{2\pi} \int \frac{d\phi _2}{2\pi} e^{-i\frac{N}{2}(\phi _2 -\phi _1)}  \times \nonumber \\
	&\times  \prod_{\mathbf{k}}\left(u_{\mathbf{k}}^2 + v_{\mathbf{k}}^2  e^{i(\phi _2 -\phi _1)} \right)	~ ,	
\end{align}
\noindent since
\begin{align*}
\left<0\right|\hat{p}_{\mathbf{k}}\left|0\right> &= \left<0\right|\hat{p}_{\mathbf{k}}^{\dagger}\left|0\right> = 0 \\
\left<0\right|\hat{p}_{\mathbf{k}}\hat{p}_{\mathbf{k}}^{\dagger}\left|0\right> &= 1	~ .
\end{align*}

\noindent We perform the change of variables:
\begin{align}
\left\{\begin{array}{ll}
    \varphi = \phi _2 - \phi _1 &  \\
    \psi = \phi _2 + \phi _1 & 
\end{array}
\right. ~,
\end{align}
which translates to a rotation of the initial integration domain and a scaling up by a factor of $2$. With the new variables the integral becomes
\begin{align}
&\left<\psi _N| \psi _N\right> = \nonumber  \\ 
	&= \frac{1}{2\left| C\right|^2} \bigg[\int_{-2\pi}^0 \frac{d\varphi}{2\pi} \int_{-\varphi}^{4\pi +\varphi} \frac{d\psi}{2\pi} e^{-i\frac{N}{2}\varphi} \prod_{\mathbf{k}}\left(u_{\mathbf{k}}^2 + v_{\mathbf{k}}^2  e^{i\varphi} \right) + \nonumber \\
	&\quad\quad + \int_{0}^{2\pi} \frac{d\varphi}{2\pi} \int_{\varphi}^{4\pi -\varphi} \frac{d\psi}{2\pi} e^{-i\frac{N}{2}\varphi} \prod_{\mathbf{k}}\left(u_{\mathbf{k}}^2 + v_{\mathbf{k}}^2  e^{i\varphi} \right)\bigg]  \nonumber
\end{align}
\begin{align}
	&= \frac{1}{4\pi ^2 \left| C\right|^2} \bigg[ \int_{0}^{2\pi} d\omega \omega e^{-i\frac{N}{2}(\omega -2\pi)} \prod_{\mathbf{k}}\left(u_{\mathbf{k}}^2 + v_{\mathbf{k}}^2  e^{i(\omega-2\pi)} \right) + \nonumber \\
	&\quad\quad +  \int_{0}^{2\pi} d\varphi (2\pi-\varphi) e^{-i\frac{N}{2}\varphi} \prod_{\mathbf{k}}\left(u_{\mathbf{k}}^2 + v_{\mathbf{k}}^2  e^{i\varphi} \right)\bigg] ~ .
\end{align}
where we performed the following additional change of variables in the first integral:
\begin{align}
&\omega \equiv 2\pi +\varphi ~ .
\end{align}
At this point we observe that only terms even in $N$ will show up in the expansion of Eq.~(\ref{eq:expansion}). That means that $N/2$ in the exponents above is an integer. That is a consequence of the fact that the product in Eq.~(\ref{eq:groundstateBCS}) essentially adds pairs of $\mathbf{k}$ states with a probability amplitude of $v_{\mathbf{k}}$ and, because of that, this state can only describe systems with an even number of particles. Using the fact that $N/2 \in \mathcal{Z}$ to simplify the first integral and renaming the dummy variable of the first integral from $\omega$ back to $\varphi$ we get:
\pagebreak[4]
\begin{align}
&\left<\psi _N| \psi _N\right> =\nonumber \\
	&= \frac{1}{4\pi ^2 \left| C\right|^2} \bigg[ \int_{-2\pi}^0 d\varphi \varphi e^{-i\frac{N}{2}\varphi} \prod_{\mathbf{k}}\left(u_{\mathbf{k}}^2 + v_{\mathbf{k}}^2  e^{i\varphi} \right) + \nonumber  \\
	&\quad +  \int_{0}^{2\pi} d\varphi (2\pi-\varphi) e^{-i\frac{N}{2}\varphi} \prod_{\mathbf{k}}\left(u_{\mathbf{k}}^2 + v_{\mathbf{k}}^2  e^{i\varphi} \right)\bigg]  \nonumber  \\ 
	&=\frac{1}{\left| C\right|^2} \int_0^{2\pi} \frac{d\varphi}{2\pi} e^{-i\frac{N}{2}\varphi} \prod_{\mathbf{k}}\left(u_{\mathbf{k}}^2 + v_{\mathbf{k}}^2  e^{i\varphi} \right)	~ .
\end{align}
Finally, demanding from $\left|\psi _N\right>$ to be normalized we find
\begin{align}
1&=\frac{1}{\left| C\right|^2} \int_0^{2\pi} \frac{d\varphi}{2\pi} e^{-i\frac{N}{2}\varphi} \prod_{\mathbf{k}}\left(u_{\mathbf{k}}^2 + v_{\mathbf{k}}^2  e^{i\varphi} \right) \Leftrightarrow \\
\left| C \right|^2 &= \int_0^{2\pi} \frac{d\varphi}{2\pi} e^{-i\frac{N}{2}\varphi} \prod_{\mathbf{k}}\left(u_{\mathbf{k}}^2 + v_{\mathbf{k}}^2  e^{i\varphi} \right) \equiv R_0^0 ~ ,	\label{eq:Lambda} 
\end{align}
where we have identified the last expression as one of the residuum integrals defined as:
\begin{align}
R_n^m & (\mathbf{k}_1 \mathbf{k}_2 \dots \mathbf{k}_m) \equiv \nonumber \\
 &\equiv\frac{1}{2\pi i} \oint dz z^{-(\frac{N}{2}-n)-1} \prod _{\mathbf{k} \neq \mathbf{k}_1, \mathbf{k} _2, \dots \mathbf{k} _m} \left( u_{\mathbf{k}}^2 + z v_{\mathbf{k}}^2\right) 	\nonumber \\
	&= \int _0^{2\pi} \frac{d\phi}{2\pi} e^{-i(\frac{N}{2}-n)\phi} \prod _{\mathbf{k} \neq \mathbf{k}_1, \mathbf{k} _2, \dots \mathbf{k} _m} \left( u_{\mathbf{k}}^2 + e^{i\phi} v_{\mathbf{k}}^2\right)	~ .
	\label{eq:App:ResInt}
\end{align}
These show up in many expectation values in the PBCS theory and they can be shown to be related to the ways that one can arrange pairs on momentum states $\mathbf{k}$.
 
 \section{The odd-even staggering formulae}
 \label{App:OES}
 
 The OES formulae for the gap $\Delta (N_0)$ are derived by taking a linear combination of values of $E(N)$ for $N$ around $N_0$:
\begin{align}
B_M = \sum _{n=-M}^{M} \alpha _{n} E(N_0+n)	~ ,		\label{eq:LinearCombination}
\end{align} 
and requiring that the contributions of $E_0(N)$ and its first $2M-1$ derivatives in $B$ vanish and that the gap $D(N)$ varies only slowly with $N$. The quantity  $E_0(N)$ is the energy of a fully paired BCS wave function and $D(N)$ is the gap defined in Eq.~(\ref{eq:DN}). The relation of $E_0(N)$ and $D(N)$ to $E(N)$ can be found in Eq.~(\ref{eq:Taylor}). Using Eq.~(\ref{eq:Taylor}) in Eq.~(\ref{eq:LinearCombination}) we get: 
\pagebreak[4]
\begin{alignat}{2}
B_M &= \sum _{n=-M}^M && \alpha _n \left[ \sum _{m=0}^\infty \frac{1}{m!}E_0^{(m)}(N_0)n^m + D(N_0+n)\right] \\
	&= \sum _{m=0}^\infty && \frac{E_0^{(m)}(N_0)}{m!} \sum _{n=-M}^M \alpha _n n^m + \nonumber \\
	& &&+ \sum_{n=-M}^M \alpha _n D(N_0+n) \nonumber  \\
	&= \sum _{m=0}^\infty && \frac{E_0^{(m)}(N_0)}{m!} \sum _{n=-M}^M \alpha _n n^m + \nonumber \\
	& &&+ \Delta (N_0)\sum_{n=-M}^M \alpha _n \frac{1-(-1)^{N_0+n}}{2}  	
\end{alignat}
where in the last step we incorporated Eq.~(\ref{eq:DN}) and assumed that $\Delta(N)$ varies slowly to pull $\Delta(N_0)$ out of the sum. As described above, demanding that the contributions of $E_0(N_0)$ and its derivatives vanish in $B_M$ we get
\begin{align}
\sum _{n=-M}^M \alpha_n n^m = 0 ~, \quad m=0,1,\dots , 2M-1~, \label{eq:OES:constr1}
\end{align}
and
\begin{align}
\sum_{n=-M}^M \alpha _n\frac{1-(-1)^{N_0+n}}{2} = 1 ~. \label{eq:OES:constr2}
\end{align}
Concerning the range of $m$ in Eq.~(\ref{eq:OES:constr1}), for a finite $M$ we have $2M+1$ coefficients $a_n$ to determine. This means that for a given finite $M$, we can demand the cancellation of the contributions from $E_0(N_0)$ and its first $2M-1$ derivatives (the $(2M+1)$-th relation comes from Eq.~(\ref{eq:OES:constr2})). Finally, for $m=0$ Eq.~(\ref{eq:OES:constr1}) is to be understood as the direct sum of the coefficients $\alpha _n$. Solving Eqs.~(\ref{eq:OES:constr1})~{\&}~(\ref{eq:OES:constr2}) one can construct an OES formula observing that, using the $\alpha _n$ coefficients found, the quantity $B_M$ simplifies to $\Delta(N_0)$,
\begin{align}
B_M = \sum _{n=-M}^{M} \alpha _{n} E(N_0+n) = \Delta (N_0) ~.   \label{eq:BMfinal}
\end{align}
For different values of $M$, Eq.~{(\ref{eq:BMfinal})} yields OES formulae of different orders.

\end{document}